\documentclass[oneside,11pt]{article}
\usepackage{amsfonts,amsmath,amssymb,amsthm}
\usepackage[colorlinks=true,linkcolor=blue,citecolor=blue,urlcolor=magenta,bookmarksopen=true,pdfstartview=FitB]{hyperref}
\usepackage{natbib}%[numbers]
\usepackage{float,graphicx,color}
\usepackage{pdfpages,multirow}
\usepackage{footmisc}
\usepackage[left=1.1in,right=1.1in,top=1.1in,bottom=1.1in]{geometry}
\usepackage[onehalfspacing]{setspace}
\usepackage{algorithm,algpseudocode}
\usepackage{hyphenat}
\newtheorem{theorem}{Theorem}[]

\newtheorem{lemma}{Lemma}[section]

\newcommand{\rbk}[1]{\left(#1\right)}
\newcommand{\sbk}[1]{\left[#1\right]}
\newcommand{\cbk}[1]{\left\{#1\right\}}

\newcommand{\bb}{\mathbb}

\newcommand{\npinull}{\hat{\pi}_{0}^{G}}

\begin{document}

\title{\vspace{-1.5cm}Multiple testing with discrete data: proportion of true null hypotheses and two adaptive FDR procedures}
\author{Xiongzhi Chen\footnote{Corresponding author: Department of Mathematics and Statistics, Washington State University,
Pullman, WA 99164, USA; Email: \texttt{xiongzhi.chen@wsu.edu}.}
, \  Rebecca W. Doerge\footnote{Office of the Dean, Mellon College of Science, 4400 Fifth Avenue, Pittsburgh, PA 15213, USA; Email: \texttt{rwdoerge@andrew.cmu.edu}.}
\ and
Joseph F. Heyse\footnote{Methodology Research, Merck Research Laboratories, 351 North Sumneytown Pike, North Wales, PA 19454, USA; Email: \texttt{joseph\_heyse@merck.com}.}
}
\date{}
\maketitle

\begin{abstract}
We consider multiple testing with false discovery rate (FDR) control when p-values have discrete and heterogeneous null distributions. We propose a new estimator of the proportion of true null hypotheses and demonstrate that it is less upwardly biased than Storey's estimator and two other estimators. The new estimator induces two adaptive procedures, i.e., an adaptive Benjamini-Hochberg (BH) procedure and an adaptive Benjamini-Hochberg-Heyse (BHH) procedure. We prove that the the adaptive BH procedure is conservative non-asymptotically. Through simulation studies, we show that these procedures are usually more powerful than their non-adaptive counterparts and that the adaptive BHH procedure is usually more powerful than the adaptive BH procedure and a procedure based on randomized p-value. The adaptive procedures are applied to a study of HIV vaccine efficacy, where they identify more differentially polymorphic positions than the BH procedure at the same FDR level.

\medskip

\end{abstract}

%%%%%%%%%%%%%%%%%%%%%%%%%%%%%%%%%%%%%%%%%%%%%%%%%%%%%%%%%%%%%%%%%%
%%%      Section: Introduction          %%%%%%%%%%%%%%%%%
%%%%%%%%%%%%%%%%%%%%%%%%%%%%%%%%%%%%%%%%%%%%%%%%%%%%%%%%%%%%%%%%%%
\section{Introduction}\label{secIntro}

Multiple testing with false discovery rate (FDR) control has been widely conducted in genomics, genetics and finance. Accordingly, many FDR procedures have been developed; see, e.g., the Benjamini-Hochberg (BH) procedure in \cite{Benjamini:1995} and Storey's procedure in \cite{Storey:2004}. However, most of these procedures were originally developed
for the ``continuous paradigm" where p-values have continuous and identical null distributions.
In contrast to the continuous paradigm, there are many multiple testing scenarios, which we refer to as the ``discrete paradigm",
where p-values have discrete and heterogeneous distributions. For example, discrete data in the form of counts have been collected in genomics using next generation sequencing (NGS) technologies \citep{Auer:2010}, in clinical studies \citep{Koch:1990}, on adverse drug reactions by the Medicines and Healthcare Products Regulatory Agency in UK, in genetics \citep{Gilbert:2005}, and in linkage disequilibrium studies \citep{Chakraborty:1987}. To analyze these data, binomial test and Fisher's exact test have been used, and their p-values have discrete and heterogeneous distributions under the null hypotheses. This leads to multiple testing in the discrete paradigm.

There has been evidence that the BH procedure and Storey's procedure tend to be less powerful or may yield unreliable results when applied to the discrete paradigm; see, e.g., \cite{Gilbert:2005} and \cite{Pounds:2006}. To develop better FDR procedures for the discrete paradigm, three major approaches have been taken: (i) modify the step-up sequence in the BH procedure according to the achievable significance level of a discrete p-value distribution; see, e.g., \cite{Tarone:1990}, \cite{Gilbert:2005} and \cite{Heyse:2011}; (ii) use randomized p-values or midP-values; see, e.g., \cite{Kulinskaya:2009}, \cite{Heller:2012} and \cite{Habiger:2015}; (iii) propose less conservative estimators of the proportion $\pi_0$ of true null hypotheses and use them to induce more powerful adaptive FDR procedures; see, e.g., \cite{Benjamini:2006}, \cite{Pounds:2006}, \cite{Blanchard:2009}, \cite{Chen:2012GenEst}, \cite{Liang:2015} and \cite{Dialsingh:2015}.

In this article, we focus on the third approach. Specifically, we propose a new estimator of $\pi_0$ for the discrete paradigm where the p-values are discrete and have heterogeneous null distributions. We prove that the new estimator is conservative and demonstrate that it is less upwardly biased than the estimators of $\pi_0$ in \cite{Storey:2004}, \cite{Benjamini:2006} and \cite{Pounds:2006}.
The new estimator induces an adaptive version of the Benjamini-Hochberg-Heyse (BHH) procedure in \cite{Heyse:2011}, referred to as the ``adaptive BHH procedure", and an adaptive version of the BH procedure, referred to as the ``adaptive BH procedure''. We prove that the
adaptive BH procedure is conservative.
Further, we empirically show that the adaptive BHH procedure is conservative and more powerful than the BHH procedure, the procedure in \cite{Habiger:2015}, the adaptive BH procedure, and the BH procedure for multiple testing based on p-values of the binomial test and Fisher's exact test.

The rest of the article is organized as follows. In \autoref{SecGenEstimator} we present the new estimator and prove its conservativeness. In \autoref{sec:adaptiveProc}, we provide the induced adaptive procedures, prove the conservativeness of the adaptive BH procedure, and discuss how to choose the guiding values for the new estimator. A simulation study for the new estimator and adaptive procedures is provided in \autoref{SecSimuStudy}. In \autoref{sec:app} we illustrate the improvement the new estimator and induced adaptive procedures can bring by applying them to a study on the efficacy of an HIV vaccine. We end the article with a discussion in \autoref{SecDiscussion}.
The proofs are relegated into the appendices.

An \textsf{R} package ``fdrDiscreteNull'' has been created to implement the new estimator and adaptive procedures, and it is available on \href{https://cran.r-project.org/}{CRAN}.

%%%%%%%%%%%%%%%%%%%%%%%%%%%%%%%%%%%%%%%%%%%%%%%%%%%%%%%%%%%%%%%%%%
%%%      section: generalized estimator of pi          %%%%%%%%%%%%%%%%%
%%%%%%%%%%%%%%%%%%%%%%%%%%%%%%%%%%%%%%%%%%%%%%%%%%%%%%%%%%%%%%%%%%
\section{A new conservative estimator of the proportion}\label{SecGenEstimator}

We start by describing a typical setting for multiple testing.
Let there be $m$ null hypotheses to test simultaneously, $I_{0}$ denote the set of true null hypotheses, and $I_{1}$ that of the
false null hypotheses. Then the proportion $\pi_0$ of true null hypotheses is just the ratio of the cardinality $m_0$ of $I_0$ to $m$, i.e., $\pi_0 = m_0 m^{-1}$.
Since $\pi_0$ is unknown and is often less than $1$, employing a good estimator of $\pi_0$ can induce an adaptive FDR procedure that is more powerful than its non-adaptive counterpart; see, e.g., \cite{Benjamini:2006} or \cite{Blanchard:2009} for examples of adaptive FDR procedures and their constructions.

It is widely known that a conservative estimator $\hat{\pi}_0$ of $\pi_0$, i.e., $\hat{\pi}_0$ having nonnegative bias, may help make its induced adaptive FDR procedure conservative. However, excessive conservativeness of $\hat{\pi}_0$ does not help increase the power of
the induced adaptive FDR procedure since it tends to reduce the magnitude of the threshold sequence of the procedure.
Further, the conservativeness of an adaptive FDR procedure can be achieved without necessarily requiring the employed estimator $\hat{\pi}_0$ to be conservative.
These facts can be seen from Sections 3 and 5 of \cite{Benjamini:2006} and Section 3 of \cite{Blanchard:2009}.
Among various estimators of $\pi_0$ (some of which have been mentioned in \autoref{secIntro}), Storey's estimator in \cite{Storey:2004} may be the most popular.
However, it is mainly designed for multiple testing in the continuous paradigm, and we will show that it can be too conservative when applied to discrete p-values. This serves as a motivation for us to develop the new estimator of $\pi_0$.

To present the results, we introduce some notations.
Assume that all p-values $\left\{  P_{i}\right\}  _{i=1}^{m}$ are defined on
the same probability space $\left(  \Omega,\mathcal{F},\mathbb{P}\right)  $,
where $\Omega$ is the sample space, $\mathcal{F}$ the $\sigma$-algebra on
$\Omega$ and $\mathbb{P}$ the probability measure.
For each $i=1,\ldots,m$, let $P_i$ be the p-value associated with the $i$th null hypothesis. For a p-value $P_i$ whose associated null hypothesis is true, let $F_i$ denote its cumulative distribution function (CDF), for
which we take the convention that any CDF is right continuous with left limits. Let $\mathsf{Unif}\left( 0,1\right) $ denote the random variable that is uniformly distributed on the closed interval $\sbk{0,1}$ and also its CDF. We assume the following:
\begin{itemize}
 % \item[A0)] Any CDF is right continuous with left limits, i.e., c\`{a}dl\`{a}g.
  \item[A0)] Each $F_{i}$ is has a non-empty support $S_{i}=\left\{ t\in \mathbb{R}:  F_{i}\left( t\right) - F_{i}\left( t-\right) >0\right\}.$
  \item[A1)] A p-value $P_{i}$ whose associated null hypothesis is true stochastically dominates $\mathsf{Unif}\left( 0,1\right) $, i.e., $F_{i}\rbk{t} \leq t$ for all $t \in \sbk{0,1}$.
\end{itemize}
We make three remarks: (i) A0) simply means that we are considering discrete p-values; (ii) A1) is a convention used in hypothesis testing;
(iii) $F_i\rbk{c} = c$ for each $c \in S_i$ for each $i=1,\ldots,m$.

%%%%%%%%%%%%%%%%%%%%%%%%%%%%%%%%%%%%%%%%%%%%%%%%%%%%%%%%%%%%%%%%%%
%%%      Section: excessive biases          %%%%%%%%%%%%%%%%%
%%%%%%%%%%%%%%%%%%%%%%%%%%%%%%%%%%%%%%%%%%%%%%%%%%%%%%%%%%%%%%%%%%
\subsection{Excessive upward bias of Storey's estimator in discrete paradigm}\label{SecExBias}

For a p-value $P_{j}$ whose associated null hypothesis is false, let $G_{j}$ be its CDF. Storey's estimator of $\pi_0$ (see Section 2.2 of \cite{Storey:2004}) is defined as
\begin{equation}\label{4A}
\hat{\pi}_{0}^{S}\left( \lambda \right) =\left( 1-\lambda \right)
^{-1}m^{-1} \left( 1+ \sum\limits_{i=1}^{m}\mathbf{1}_{\left\{ P_{i}>\lambda \right\} } \right)
\end{equation}
for a tuning parameter $\lambda \in [0,1)$, where $\mathbf{1}_{A}$ is the indicator function of the set $A$. Its bias is the sum of
$\left( 1-\lambda \right)^{-1}m^{-1}$ and
\begin{equation}\label{eqBiasNullStorey}
b_0 = \left(
1-\lambda \right) ^{-1}m^{-1} \sum\nolimits_{i\in I_{0}}\sbk{ \lambda-F_{i}\left( \lambda \right)}
\end{equation}
and
\begin{equation}\label{biasStorey2}
b_1 = \left(
1-\lambda \right) ^{-1}m^{-1}\sum\nolimits_{i\in I_{1}}\sbk{ 1-G_{i}\left( \lambda\right) }.
\end{equation}

Call a p-value whose associated null hypothesis is true a ``null p-value" and that whose associated null hypothesis is false an ``alternative p-value". Then the bias $b_0$ is associated with the null p-values, and it is zero when they are uniformly distributed on $[0,1]$. However, $b_0$ is usually positive when p-values have discrete distributions with different supports. In contrast, the bias associated with the alternative p-values, $b_1$, is usually positive regardless of if the p-values have continuous or discrete distributions, and it cannot be reduced unless information on the p-value distributions under the alternative is available.
Fortunately, when p-values have discrete distributions, it is possible to significantly reduce the bias $b_0$ by choosing for each p-value its own tuning value from the support of its CDF. This is achieved by the new estimator to be presented next.

\subsection{New estimator and its conservativeness}\label{sec:newestpi0}

The new estimator, denoted by $\hat{\pi}_{0}^{G}$, of the proportion of true null hypotheses is stated in \autoref{Alg:newest}.
To explain the rationale behind $\hat{\pi}_{0}^{G}$, we start from a trial estimator $\beta\left( \tau _{j}\right)$ for a fixed $j$ stated in \eqref{eqNewEstIter}. There are $3$ components in $\beta\left( \tau _{j}\right)$, each with its own functionality:
\begin{itemize}
  \item  The first summand $\left(\left(1-\tau_j\right)m\right)^{-1}$ in \eqref{eqNewEstIter} is technical and used to prove the conservativeness of the adaptive BH procedure in \autoref{ThmConservABH}. When $\tau_j$ is small and $m$ is large, this term is negligible.

  \item  The second summand in \eqref{eqNewEstIter} is the key component and specifically designed for discrete p-values. Note that $\lambda_{ij}$ is chosen from the support $S_i$ of the CDF of p-value $P_i$ for $1 \le i \le m$. So, for each $i \in I_{0}$, the term $\lambda - F_i\left(\lambda\right)$ in the expression for the bias $b_0$ in \eqref{eqBiasNullStorey} for Storey's estimator $\hat{\pi}_0^S$ changes into $\lambda_{ij} - F_i\left(\lambda_{ij}\right)$, being exactly $0$. Therefore, the bias of $\beta \left( \tau _{j}\right)$ associated with the null p-values is $0$; see \autoref{thm:Conservative} for a justification.

  \item The third summand in \eqref{eqNewEstIter} is a deterministic quantity and specifically designed for discrete p-values. It accounts a null hypothesis as being true if the support of the CDF of its associated p-value is a singleton or equivalently if the CDF of its associated p-value is a Dirac mass. For example, when the total observed count from two independent binomial (or Poisson) random variables is $1$,  for a Fisher's exact test (or binomial test) the CDF of its two-sided p-value is a Dirac mass (our simulation study in \autoref{SecSimuStudy} will simulate such cases). This summand seems only to add to the upward bias of the new estimator. However, \autoref{thm:Conservative} shows that it is not so.
\end{itemize}

\begin{algorithm}[t!]
\begin{algorithmic}[1]
\caption{New estimator of the proportion of true null hypotheses}\label{Alg:newest}
\State
Let $Q_s=\left\{1,\ldots,s\right\}$ for each natural number $s$.
Set $q_i = \inf\left\{c: c \in S_i\right\}$ for each $i \in Q_m$ and $\gamma= \max\left\{q_i: i \in Q_m\right\}$.
Pick a sequence of $n$ increasing, equally spaced ``guiding values" $\left\{\tau_j\right\}_{j=1}^n$ such that $\tau_0 \le \tau_1 \le \ldots \le \tau_n < 1$, for which $\tau_0$ is set as follows: (i) if $\gamma=1$, set $\tau_0 = \max\left\{q_i: q_i <1\right\}$; (ii) if $\gamma<1$, set $\tau_0 = \gamma$.

\State %\parbox[t]{\dimexpr\linewidth-\algorithmicindent}{
   Let $C=\left\{i \in Q_m : q_i = 1 \right\}$. For each $i \in Q_m \setminus C$ and $j \in Q_n$, set
   \begin{equation*}%\label{eq:lambdaij}
  T_{ij}=\left\{ \lambda \in S_{i} :\lambda\leq \tau _{j} \right\}
\end{equation*}
and $\lambda _{ij}=\sup\left\{\lambda: \lambda \in T_{ij} \right\}$. For each $j \in Q_n$, define the ``trial estimator"
\begin{equation}\label{eqNewEstIter}
\beta\left(\tau_j\right)  =\frac{1}{\left(  1-\tau_{j}\right)  m}%
+\frac{1}{m}\sum\nolimits_{i \in Q_m \setminus C}\frac{\mathbf{1}_{\left\{  P_{i}%
>\lambda_{ij}\right\}  }}{1-\lambda_{ij}} + \frac{1}{m}\left\vert C \right\vert,
\end{equation}
where $\left\vert A\right\vert $ the cardinality of a set $A$. Truncate $\beta\left(\tau_j\right)$ at $1$ when it is greater than $1$.
%\strut}

\State %\parbox[t]{\dimexpr\linewidth-\algorithmicindent}{
  Set
  \begin{equation}\label{eq:newEst}
    \hat{\pi}_{0}^{G}=\frac{1}{n}\sum_{j=1}^n \beta\left(\tau_j\right)
  \end{equation}
  as the estimate of $\pi_0$.
%\strut}
\end{algorithmic}
\end{algorithm}

\begin{figure}[t!]
\centering
\includegraphics[width=.7\textwidth]{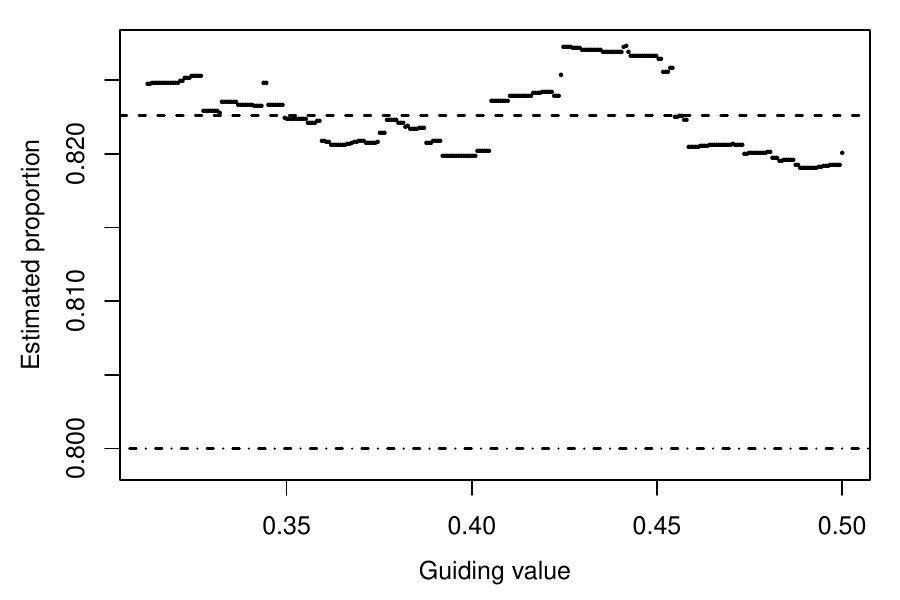}
\caption[Illustrate new est] {The new estimator applied to two-sided p-values of binomial tests. The conservative, trial estimator $\beta \left( \tau\right)$ (on the vertical axis) of the proportion $\pi_0$ is plotted against the guiding value $\tau$. The new estimator of $\pi_0$ is the average of the $\{\beta\left( \tau_j\right)\}_{j=1}^n$ for an adaptively chosen sequence of $n$ guiding values $\left\{ \tau_j \right\}_{j=1}^n$, and it is indicated by the horizontal dashed line; see \autoref{Alg:newest} for details on constructing the new estimator. In this example, the true $\pi_0 = 0.8$ (indicated by the dot dashed line), the new estimator is $0.8226$ (indicated by the dashed line), Storey's estimator in \cite{Storey:2004} with $\lambda=0.5$ is $1$,
the estimator in \cite{Benjamini:2006} based on the median of p-values is $1$, the estimator in \cite{Pounds:2006} is $0.999$.}
\label{figIllustrate}
\end{figure}

Since each $\beta\left( \tau _{j}\right)$ tends to have smaller upward bias than $\hat{\pi}_0^S$, the new estimator $\hat{\pi}_{0}^{G}$, being the average of $\left\{\beta\left( \tau _{j}\right)\right\}_{j=1}^n$, will also tend to be so and be more stable than each one of them. An illustration of the construction of $\hat{\pi}_{0}^{G}$ is provided in \autoref{figIllustrate}.
Note that $\npinull$ is a functional of the supports of all p-value CDFs and is essentially different than the estimators of $\pi_0$ in \cite{Storey:2004}, \cite{Benjamini:2006}, \cite{Pounds:2006}, \cite{Liang:2012} and \cite{Liang:2015}.

The following \autoref{thm:Conservative} shows that $\hat{\pi}_0^G$ is conservative. However, we point out again that conservativeness of $\hat{\pi}_0^G$ is not necessarily needed for its induced adaptive procedure to be conservative; see \autoref{ThmConservABH} in \autoref{sec:adaptiveProc}. We will discuss in \autoref{sec:adaptiveProc} the choice of guiding values $\left\{\tau_j\right\}_{j=1}^n$ in \autoref{Alg:newest} once we prove the conservativeness of the adaptive BH procedure.

\begin{theorem}\label{thm:Conservative}
Recall $C=\left\{i \in Q_m : q_i = 1 \right\}$, where $q_i = \inf\left\{c: c \in S_i\right\}$.
For each $1 \le j \le n$, the bias of the trial estimator $\beta\left(  \tau_{j}\right)$ is
\begin{equation}\label{eqBiasTrailA}
\delta_j = \bb{E}\left( \beta\left(  \tau_{j}\right) -\pi _{0}\right)= \frac{1}{\left(1-\tau_j\right)m}
+\frac{1}{m}\sum\nolimits_{i\in I_{1}\setminus C}\frac{1-G_{i}\left( \lambda _{ij}\right) }{1-\lambda_{ij}},
\end{equation}
and $\delta_j \ge 0$, where $\bb{E}$ denotes expectation. So, the bias of $\hat{\pi}_0^G$ is
$\delta = n^{-1}\sum_{j=1}^n \delta_j $. Therefore,
$\beta\left(  \tau_{j}\right)$ is conservative for each $1 \le j \le n$, and so is the new estimator $\hat{\pi}_{0}^{G}$.
\end{theorem}

%%%%%%%%%% illustrate of new estimator

\autoref{thm:Conservative} shows that the summand $m^{-1}\left\vert C \right\vert$ in the definition \eqref{eqNewEstIter} of the trial estimator adds $0$ bias to the new estimator.
It is hard to determine which among Storey's estimator $\hat{\pi}_0^S$ and the new estimator $\hat{\pi}_0^G$ is less conservative without information on the CDF's $G_i$ of the alternative p-values. Specifically, if $\delta \le b_0+b_1$, then $\bb{E}\left(\hat{\pi}_0^G\right) \le \bb{E}\left(\hat{\pi}_0^S\right)$, i.e., $\hat{\pi}_0^G$ is less conservative than $\hat{\pi}_0^S$, where $b_0$ is defined in \eqref{eqBiasNullStorey} and $b_1$ in \eqref{biasStorey2}. However, to show that $\delta \le b_0+b_1$ to hold, restrictive assumptions on $G_i$'s may be needed when the p-value distributions are discrete and heterogenous. So, we do not pursue this further here.

\section{Two adaptive procedures induced by the new estimator}\label{sec:adaptiveProc}

Now we introduce two adaptive FDR procedures based on the new estimator $\hat{\pi}_0^G$, i.e., the adaptive BH procedure and the adaptive BHH procedure. Let the nominal FDR level be $\alpha \in \left(0,1\right)$. The adaptive BH procedure is obtained by applying the BH procedure in \cite{Benjamini:1995} at new nominal FDR level $\alpha / \hat{\pi}_0^G$.
Similarly, the adaptive BHH procedure is obtained by applying the BHH procedure in \cite{Heyse:2011} at new nominal FDR level $\alpha / \hat{\pi}_0^G$. For readers' convenience, the BH procedure and the BHH procedure are provided in \autoref{secBHHProc}, and two misinterpretations of the BHH procedure
are given in \autoref{secMinsiterp}. Note that the BHH procedure accounts for the discreteness of p-value distributions, can be regarded as an extension of the BH procedure, and has been shown to be more powerful than the BH procedure under some settings; see \cite{Heyse:2011} for a simulation study on this.
The adaptive procedures induced by $\hat{\pi}_0^G$ can be more powerful than their nonadaptive counterparts when $\pi_0 <1$ and $\hat{\pi}_0^G \neq 1$.

\subsection{Conservativeness of the adaptive BH procedure}

To state the result on the conservativeness of the adaptive BH procedure, we introduce some notations.
Recall that $I_0$ is the index set of true null hypotheses. For each $k\in I_{0}$, let%
\[
\mathbf{p}_{0,k}=\left\{  P_{1},\ldots,P_{k-1},0,P_{k+1},\ldots,P_{m}\right\}  .
\]
Correspondingly, let $\beta_{k}\left(\tau_j\right)  $ be the trial
estimator with guiding value $\tau_{j}$ obtained by applying \autoref{Alg:newest} to
$\mathbf{p}_{0,k}$ and set
\begin{equation}
\hat{\pi}_{0,k}^{G}=n^{-1}\sum_{j=1}^{n}\beta_{k}\left(\tau_j\right)
.\label{eqe3}%
\end{equation}
Recall $Q_s = \left\{1,\ldots,s\right\}$ for any natural number $s$ and $C=\left\{i \in Q_m : q_i = 1 \right\}$ where $q_i = \inf\left\{c: c \in S_i\right\}$.

\begin{theorem}\label{ThmConservABH}
If $I_0 \subseteq Q_m\setminus C$ and the p-values are independent, then the following hold:
\begin{enumerate}
  \item For each $k \in I_0$ and $j \in Q_n$ with any positive integer $n$,
\begin{equation}\label{eqe2A}
\mathbb{E}\left( 1/\beta_{k}\left(\tau_j\right) \right) \leq \pi_{0}^{-1}
\quad \text{and} \quad
\mathbb{E}\left(  1/\hat{\pi}_{0,k}^{G} \right)  \leq \pi_{0}^{-1}.
\end{equation}

\item The adaptive BH procedure induced by the new estimator $\hat{\pi}_{0}^{G}$ is conservative.
\end{enumerate}

\end{theorem}

\autoref{ThmConservABH} justifies a crucial property, i.e., inequality \eqref{eqe2A}, that may be used to prove the conservativeness of the adaptive BHH procedure. Further, it ensures that the adaptive BH procedure is conservative and potentially more powerful than the BH procedure when $\pi_0 < 1$ and $\hat{\pi}_0^G < 1$.
The condition $I_0 \subseteq Q_m\setminus C$ requires that no null p-value should have its CDF as a Dirac mass, which easily holds for binomial test (or Fisher's exact test) as long as the total observed count is bigger than $1$ for each pair of independent Poisson (or binomial) random variables. Even though this condition is violated in our simulation study (see \autoref{SecSimuStudy}), our simulation results show that the adaptive BH procedure is still conservative when applied to independent p-values.

\subsection{Adaptive choice of guiding values for the new estimator}\label{sec:conserve}

In this section we discuss the choice of the guiding values $\left\{\tau_j\right\}_{j=1}^n$ in \autoref{Alg:newest}.
Based on the decomposition of the bias of the new estimator $\hat{\pi}_{0}^{G}$ given in the proof of \autoref{thm:Conservative}, it is better to pick a $\tau_n$, the maximum of the $\tau_j$'s, that is much smaller than $1$, so that the term $\left(\left(  1-\tau_{j}\right)  m\right)^{-1}$ is negligible when $m$ is relatively large.
On the other hand, $\max_{i \in Q_m\setminus C} \lambda_{ij} \le \tau_j$ for each $j \in Q_n$, and a null p-value tends to assume relatively large values. So, if the CDF's $G_i$ of the alternative p-values increase much slower than the identity function, a small $\tau_j$ may make big the second summand in the definition of $\beta\left(\tau_j\right)$, leading to large upward bias of $\hat{\pi}_{0}^{G}$ and smaller gain in power for the induced adaptive FDR procedure. Thus, our principle is not to set $\tau_1$, the smallest of the $\tau_j$'s, too small and not to set $\tau_n$ big.

Specifically, the guiding values $\left\{\tau_j\right\}_{j=1}^n$ are set as follows.
Recall $\tau_{0}$ defined in \autoref{Alg:newest}. If $\tau_{0}<0.5$, set $\tau_{1}=
\tau_{0}+0.5\times\left(  0.5-\tau_{0}\right)    $, $n=100$ and $\tau_{n}=0.5$, meaning that
the step size $d=\tau_{j+1}-\tau_{j}=100^{-1}\left(  \tau
_{n}-\tau_{1}\right)  $; otherwise, set $\tau_{1}=\tau_{n}=0.5$ and $n=1$.
In other words, when $\tau_{0}<0.5$, only $100$ trial estimators will be computed, so as not to take much computational time. Note that \autoref{thm:Conservative} and \autoref{ThmConservABH} are valid for any guiding sequence described in \autoref{Alg:newest}.
Our simulation study in \autoref{SecSimuStudy} will show that the above choice for $\left\{\tau_j\right\}_{j=1}^n$ works well and maintains the accuracy and stability of the new estimator and the conservativeness of the induced adaptive procedures.

%%%%%%%%%%%%%%%%%%%%%%%%%%%%%%%%%%%%%%%%%%%%%%%%%%%%%%%%%%%%%%%%%%%%%%%%%%%%%
%%%%%%%%%%%%  Simulation stuides                        %%%%%%%%%%%%
%%%%%%%%%%%%%%%%%%%%%%%%%%%%%%%%%%%%%%%%%%%%%%%%%%%%%%%%%%%%%%%%%%%%%%%%%%%%%
\section{Simulation study}\label{SecSimuStudy}

Now we assess the performance of the new estimator $\hat{\pi}_{0}^{G}$ and adaptive procedures
via simulation studies based on discrete p-values of binomial tests and Fisher's exact tests (FETs).
The estimators of $\pi _{0}$ we compare are the new estimator $\hat{\pi}_{0}^{G}$, Storey's estimator
$\hat{\pi}_{0}^{S}\left( \lambda \right)$ in \eqref{4A} with $\lambda =0.5$, the estimator
%\begin{equation*}%\label{eqPndestA}
$\hat{\pi}_{0}^{PC} = \min \cbk{ 1,2m^{-1}\sum\nolimits_{i=1}^{m}P_{i}}$
%\end{equation*}
in \cite{Pounds:2006}, and the median based estimator
%\begin{equation*}%\label{eqBenest}
$\hat{\pi}_{0}^{BKY}= m^{-1}\left( m-\left[ m/2\right] +1 \right) \left( 1-P_{\left( \left[ m/2\right]
\right) }\right)^{-1}$
%\end{equation*}
in \cite{Benjamini:2006}.

We choose $\hat{\pi}_{0}^{S}\left( 0.5\right) $ since other methods provided by the \textsf{qvalue} package to implement $\hat{\pi}
_{0}^{S}$ give more upwardly biased estimate of $\pi_0$ than $\hat{\pi}%
_{0}^{S}\left( 0.5\right) $ for the simulations. In contrast, we set $\hat{\pi}_{0}^{BKY}$ using the median
of the p-values to make it robust since it is not designed for discrete p-values.
However, we will not investigate the estimator
%\begin{equation}\label{eq:PndestB}
$ \hat{\pi}_{0}^{PC\ast}=\min \left\{ 1,m^{-1}\sum\nolimits_{i=1}^{m} P_{i}\mu_i^{-1}\right\}$
%\end{equation}
proposed by \cite{Pounds:2006} where $\mu_i$ is the mean of $P_i$ computed under the null hypothesis, since
we have observed in \cite{Chen:2012GenEst}
that $\hat{\pi}_{0}^{PC\ast}$ is usually $1$ when $\pi_0 \geq 0.5$ for a similar simulation setup (see \autoref{SecSimSteup} for the simulation design).
In addition, we will not consider estimators in \cite{Dialsingh:2015} since they are based on the two-groups model for the p-values.

We will compare the adaptive BHH procedure (``aBHH") with the BHH procedure in \cite{Heyse:2011}, the procedure (denoted by ``SARP") in \cite{Habiger:2015} that is based on applying Storey's procedure in \cite{Storey:2004} to randomized p-values obtained from the discrete p-values, the adaptive BH procedure (``aBH"), and the BH procedure (``BH''). However, we will not investigate the fuzzy FDR procedure in \cite{Kulinskaya:2009} or the discrete Benjamini-Liu (``DBL") procedure in \cite{Heller:2012}, since results from the former do not usually have a straightforward interpretation and the latter is not necessarily as powerful as the BHH procedure at the same nominal FDR level.
Finally, we will implement SARP exactly according to \cite{Habiger:2015}.

\subsection{Simulation study design}\label{SecSimSteup}

The simulation is set up as follows:
\begin{enumerate}
\item Set $m=1000$, $\pi _{0}\in \left\{ 0.5,0.6,0.7,0.8,0.95\right\} $, $m_0 = m\pi_0$, and nominal FDR level to be $0.05$.
For each value for $\pi _{0}$, do the following:

\item Generate data:

\begin{enumerate}
\item Poisson data: let \textsf{Pareto}$\left( l,\sigma\right) $ denote the Pareto
distribution with location $l$ and shape $\sigma$ and $\mathsf{Unif}\left(
a,b\right) $ be the uniform distribution on the interval $[a,b]$. Generate $m$ $\theta _{i1}$'s independently from
$\mathsf{Pareto}\left( 3,8\right) $. Generate $m_{1}$ $\rho _{i}$'s independently from $\mathsf{Unif}\left(
1.5,4.5\right) $. Set $\theta _{i2}=\theta _{i1}$ for $1\leq i\leq m_{0}$ but $%
\theta _{i2}=\rho _{i}\theta _{i1}$ for $m_{0}+1\leq i\leq m$. For each $1 \leq i \leq m$ and $g \in \cbk{1,2}$,
independently generate a count $\xi _{ig}$ from the Poisson distribution $\mathsf{Poisson}\left( \theta _{ig}\right) $
with mean $\theta _{ig}$.

\item Binomial data: generate $\theta_{i1}$ from $\mathsf{Unif}\left(
0.15,0.2\right)  $ for $i =1,\ldots,m_{0}$ and set $\theta_{i2}=\theta_{i1}$ for $i =1,\ldots,m_{0}$. Set $\theta_{i1}=0.2$ and
$\theta_{i2}=0.5$ for $i =m_{0}+1,\ldots,m$. Set $n=20$, and for each $g \in \{1,2\}$ and $i$, independently generate a count $\xi _{ig}$ from the binomial distribution $\mathsf{Bin}\left( \theta_{ig}, n\right)$ with probability of success $\theta_{ig}$ and number of trials $n$.

\end{enumerate}

\item With $\xi _{ig}$, $g=1,2$ for each $i$, conduct the binomial test for Poisson data and Fisher's exact test (FET) for
binomial data to test
%\begin{equation*}
$H_{i0}:\theta _{i1}=\theta _{i2}\text{ versus }H_{i1}:\theta _{i1}\neq \theta _{i2}$
%\end{equation*}
and obtain the two-sided p-value $P_{i}$ of the test, or to test
%\begin{equation*}
$H_{i0}:\theta _{i1}=\theta _{i2}\text{ versus }H_{i1}:\theta _{i1} < \theta _{i2}$
%\end{equation*}
and obtain the one-sided p-value $P_{i}$ of the test, where $H_{i0}$ denotes a true null hypothesis. Observe that
$\theta_{i1} < \theta_{i2}$ for each false null hypothesis for the simulated data.

\item Apply the four estimators of $\pi_0$ and FDR procedures to the $m$ p-values $\left\{ P_{i}\right\} _{i=1}^{m}$ or the corresponding randomized p-values.

\item Repeat Steps 2. to 4. $250$ times to obtain statistics for the performance of each estimator and FDR procedure.
\end{enumerate}

For the simulated data, the difference between the Poisson means ranges from small to large values, so that the binomial tests are not dominated by very large effect sizes and that the discrete p-values induced by these tests range more sufficiently from $0$ and $1$. The simulation scheme for the binomial data is similar to that employed by \cite{Gilbert:2005} for a study on the genetics of immunological difference to the HIV. In view of these, our simulation study design induces fair comparison between the estimators of $\pi_0$ and FDR procedures and is practical.

For each test, its two-sided p-value is computed according to the formula in \cite{Agresti:2002}, i.e., it is the
probability computed under the null hypothesis of observing values of the test statistic that are equally likely as or less likely than the observed test statistic.
For the simulated data, $\theta_{i1} < \theta_{i2}$ for each false null hypothesis. So, a one-sided p-value is directly computed as the probability under the null
hypothesis of observing values of the test statistic that are smaller than or equal to the observed test statistic.

\subsection{Simulation study results}\label{secSimRes}

\begin{figure}[htp]
\centering
\includegraphics[height=0.73\textheight,width=1\textwidth]{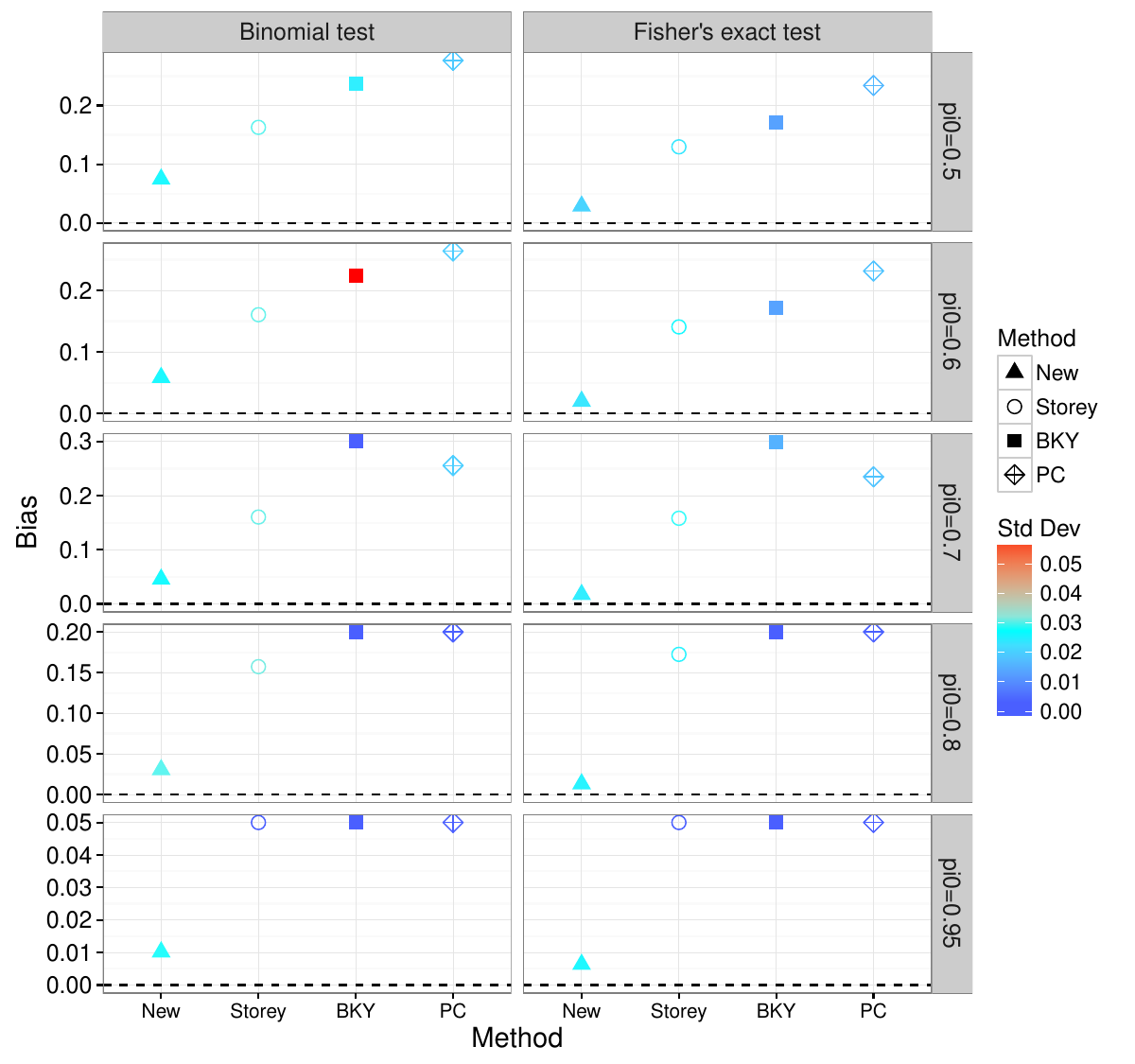}
\caption[Bias and std one-sided] {
Bias and standard deviation (indicated by the color legend ``Std Dev") of each estimator of the proportion $\pi_0$ of true null hypotheses. All estimators have been applied to one-sided p-values of a type of test indicated by the horizontal strip name. The dashed line marks zero bias; ``pi0'' the vertical strip names refers to $\pi _{0}$.
An estimator of $\pi_0$ is said to be better if it has smaller non-negative bias and small standard deviation. The new estimator (indicated by ``New'' and the triangle) is conservative and the best. An estimator can have standard deviation very close $0$ when it is always very close to $1$, and this happens to the estimators in \cite{Storey:2004}, \cite{Pounds:2006} and \cite{Benjamini:2006} when $\pi_0 = 0.8$ or $0.95$.
}
\label{figbiasstdLT}
\end{figure}

\begin{figure}[htp]
\centering
\includegraphics[height=0.73\textheight,width=1\textwidth]{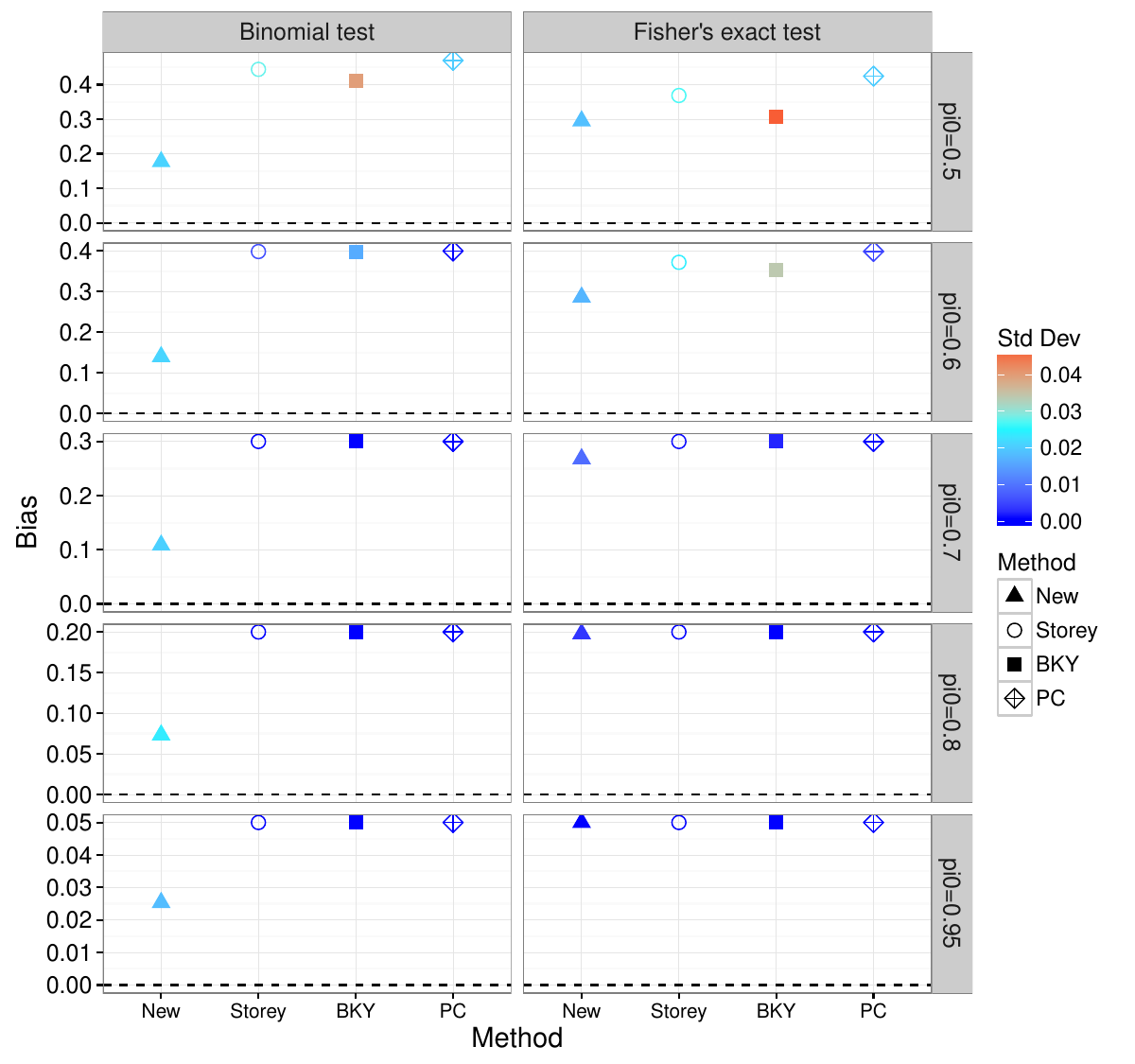}
\caption[Bias and std two-sided] {Bias and standard deviation (indicated by the color legend ``Std Dev") of each estimator of the proportion $\pi_0$ of true null hypotheses. All estimators have been applied to two-sided p-values of a type of test indicated by the horizontal strip name. The dashed line marks zero bias; ``pi0'' the vertical strip names refers to $\pi _{0}$.
An estimator of $\pi_0$ is said to be better if it has smaller non-negative bias and small standard deviation. The new estimator (indicated by ``New'' and the triangle) is conservative and the best. An estimator can have standard deviation very close $0$ when it is always very close to $1$, and this happens to the estimators in \cite{Storey:2004}, \cite{Pounds:2006} and \cite{Benjamini:2006} when $\pi_0 = 0.8$ or $0.95$.}
\label{figbiasstd}
\end{figure}

An estimator of the proportion $\pi_0$ is better if it is less conservative (i.e., having smaller upward bias), has small standard deviation, and induces a conservative adaptive FDR procedure. \autoref{figbiasstdLT} and \autoref{figbiasstd} present the biases and standard deviations of the estimators when
they are applied to p-values of binomial tests or FETs. For all settings we have considered, the new estimator $\npinull$ is conservative, the most accurate, and stable (i.e., having small standard deviation). The improvement of $\npinull$ over the other estimators can be considerable when $\pi_0$ is not very close to $1$. In contrast, all other three estimators have more upward biases than the new estimator, and they can be very close to $1$ quite often even when $\pi_0=0.8$.

All estimators tend to be slightly more conservative when applied to two-sided p-values than one-sided p-values. This is due to two things: (1) two-sided p-value are more likely to be $1$ than one-sided ones, and in the simulation the number of two-sided p-values being $1$ is often larger than that of one-sided p-values; (2) in the simulation there are p-values whose CDF's are Dirac masses at different singletons, i.e., there are p-values which take only the value $1$ almost surely, inducing more upward bias to each estimator.

\begin{figure}[htp]
\centering
\includegraphics[height=0.73\textheight,width=1\textwidth]{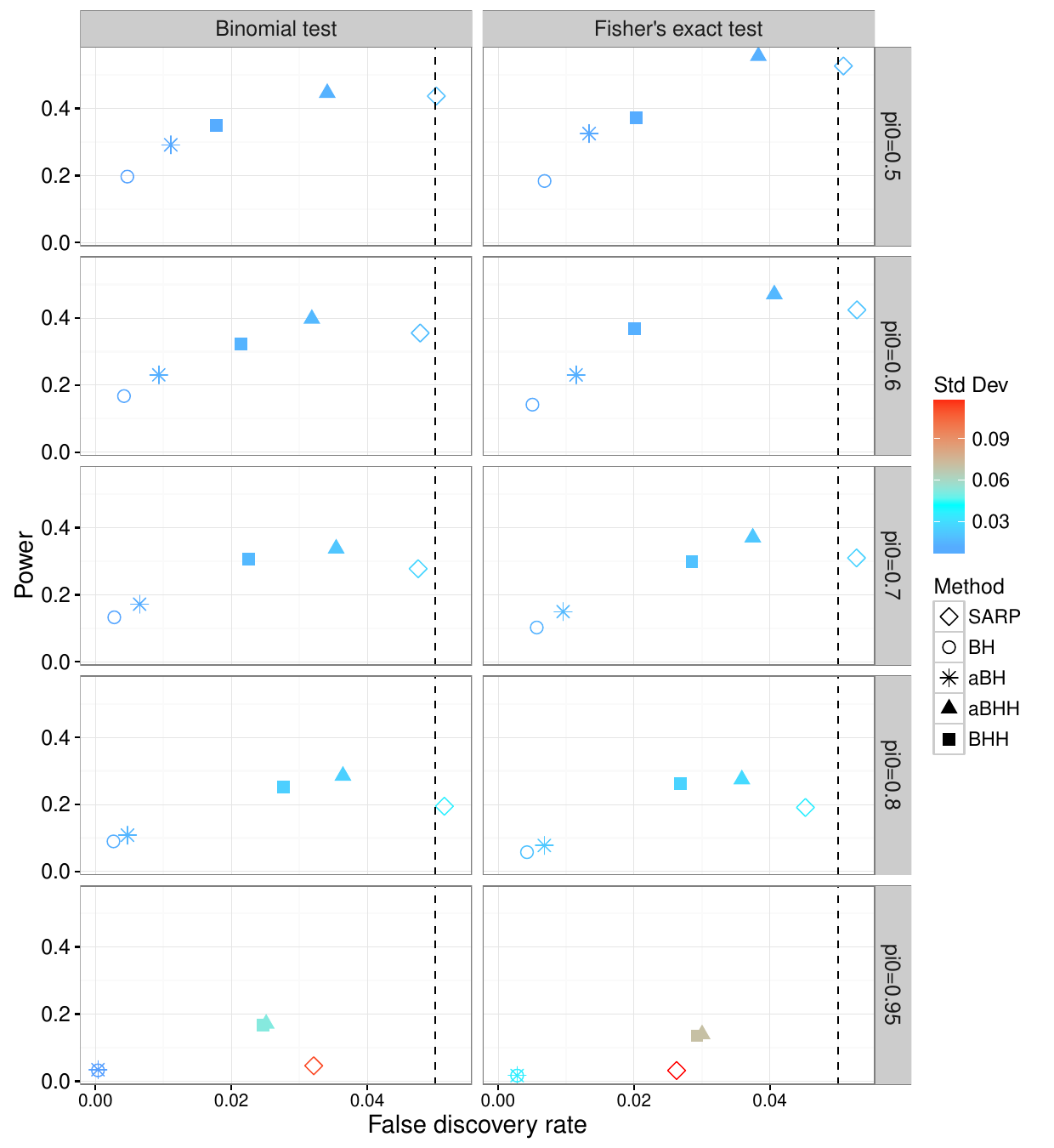}
\caption[FDR and power: lower tail] {
False discovery rate (FDR) and power of the competing FDR procedures when they are applied to one-sided p-values of a type of test indicated by the horizontal strip name. In the vertical strip names, ``pi0'' refers to $\pi _{0}$; the color gradient is the standard deviation (Std Dev) of the false discovery proportion whose expectation is the FDR.
The adaptive BHH procedure ``aBHH", indicated by solid triangle, has FDR below the nominal FDR level $0.05$, and it is the most powerful. However, the procedure ``SARP" in \cite{Habiger:2015} may have slightly larger FDRs than the nominal level in this setting. This is likely because the estimator of the proportion $\pi_0$ employed by SARP under-estimates $\pi_0$.}
\label{figPwrLower}
\end{figure}

\begin{figure}[htp]
\centering
\includegraphics[height=0.73\textheight,width=1\textwidth]{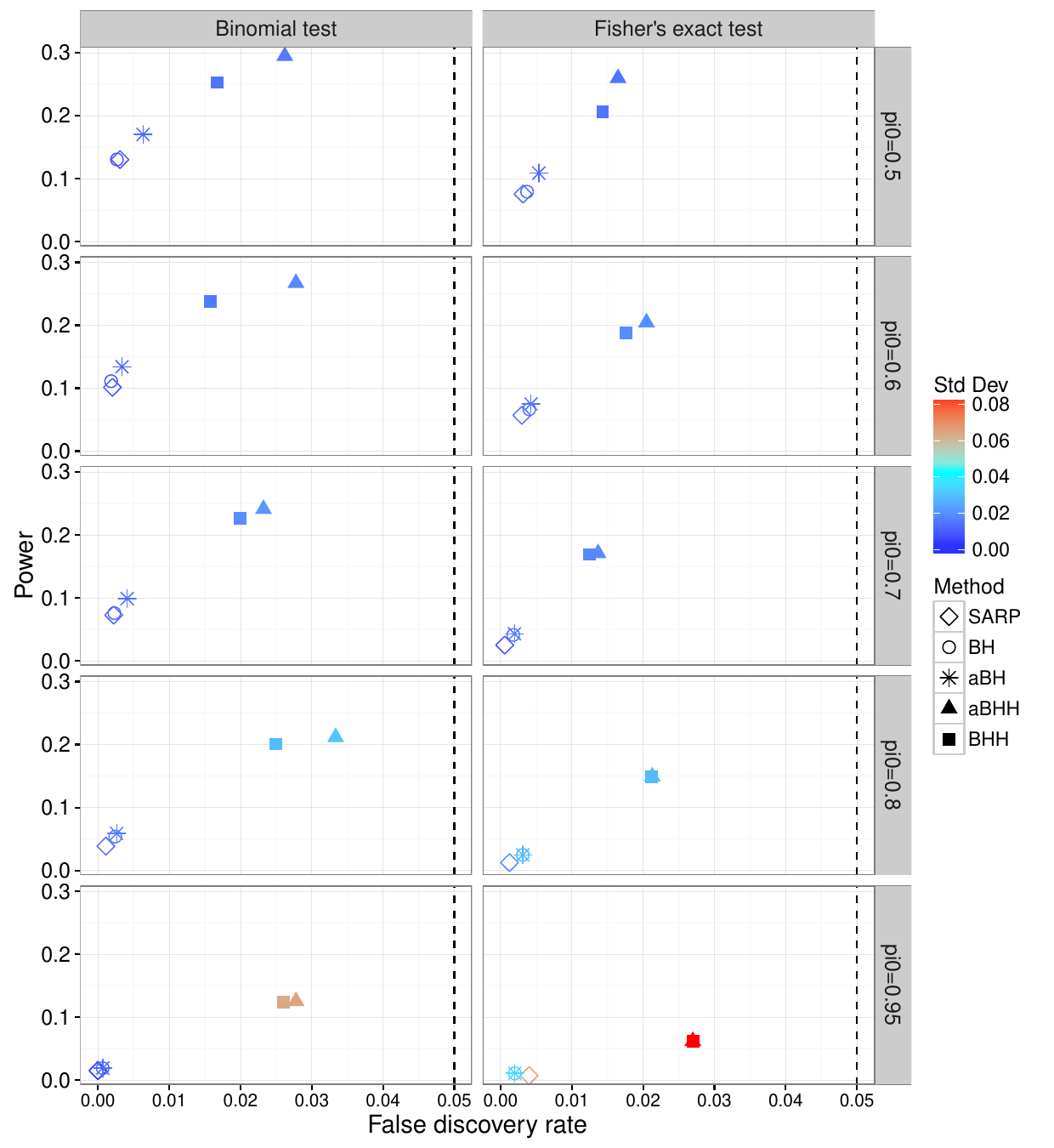}
\caption[FDR and power, two-sided] {False discovery rate (FDR) and power of the competing FDR procedures when they are applied to two-sided p-values of a type of test indicated by the horizontal strip name. In the vertical strip names ``pi0'' refers to $\pi _{0}$; the color gradient is the standard deviation (Std Dev) of the false discovery proportion whose expectation is the FDR. All procedures have FDRs below the nominal FDR level $0.05$, and the adaptive BHH procedure ``aBHH",  indicated by solid triangle, is the most powerful.}
\label{figPwrTwoside}
\end{figure}

We use the expectation of the true discovery proportion (TDP), defined as the ratio of the number of rejected false null hypotheses to the total number of false null hypotheses, to measure the power of an FDR procedure.
Recall that the FDR is the expectation of the false discovery proportion (FDP, \citealp{Genovese:2002}). We also report the standard deviations of the FDP and TDP since smaller standard deviations for these quantities mean that the corresponding procedure is more stable in FDR and power.
An FDR procedure is better if it is more powerful at the same nominal FDR level and stable. \autoref{figPwrLower} and \autoref{figPwrTwoside} record the FDRs and powers of the five FDR procedures, BH, aBH, BHH, aBHH and SARP, when they are applied to p-values of binomial tests or FETs at nominal FDR level $0.05$. The adaptive BH procedure and adaptive BHH procedure are conservative (i.e., their FDRs are upper bounded by the nominal FDR level) and stable. In particular, the adaptive BHH procedure is the most powerful among the five for all settings we have considered. This is expected since (i) the new estimator is less conservative than the other three estimators, (ii) the adaptive BHH procedure improves the BHH procedure and the latter the BH procedure, (iii) SARP constructs randomized p-values based on the observed, discrete p-values and is exactly Storey's procedure in \cite{Storey:2004} applied to the randomized p-values, and (iv) the adaptive BH procedure and Storey's procedure differ only by the estimators of $\pi_0$ they employ. However, the FDRs of the BH, aBH, BHH, and aBHH procedures are well below the nominal level when applied to two-sided p-values, indicating room for further improvement on the power of the aBHH procedure.

For two-sided p-values of FETs, the estimator of $\pi_0$ are very conservative due to the reasons described previously and the improvements of the adaptive FDR procedures upon their non-adaptive counterparts are small; see \autoref{figPwrTwoside}. For this setting, the FDR procedures are less powerful when there are more p-values taking values $1$, likely due to a potential power decrease in their associated tests when the corresponding observed total counts are small. This is more obvious when there is a considerable proportion of p-values whose CDF's are Dirac masses since these p-values almost surely are $1$ and their associated null hypotheses are usually not rejected by the FDR procedures even if some of them are false null hypotheses. However, in these situations, the BHH procedure is much more powerful than the BH procedure, adaptive BH procedure and SARP, indicating the advantage of the BHH procedure in settings where tests have low (to moderate) power or p-value CDF's are Dirac mass.

We have found that the estimates of $\pi_0$ given by SARP have relatively large variance and often are much smaller than $\pi_0$. This may explain why the FDRs of SARP are slightly larger than the nominal level, i.e., SARP being anti-conservative, when applied to one-sided p-values and $\pi_0$ is not close to $1$; see the right column of \autoref{figPwrLower}. This reveals that, due to the use of randomized p-values, SARP may introduce unfavorable randomness and instability to multiple testing in the discrete paradigm.
In addition, we have found out that the adaptive BH procedure is less powerful than SARP when applied to one-sided p-values and that they are equally powerful when applied to two-sided p-values. However, since SARP can be anti-conservative, for multiple testing based on two-sided p-values of binomial tests or FETs, it may be better to apply the adaptive BH procedure rather than SARP.

A simulation study under approximate positive, block-wise dependence is given in \autoref{secSimDep}. For each setting of this simulation indicated by a value of $\pi_0$ and a type of test, the empirical CDF of the p-values has a bimodal distribution, with well separated modes and one mode being around $0$. All estimators of $\pi_0$ are more conservative than when they are applied to independent p-values. In particular, Storey's estimator with tuning parameter $\lambda=0.5$
is more conservative than the other estimators. For one-sided p-values, the estimator $\hat{\pi}_{0}^{PC}$ in \cite{Pounds:2006} seems to be the least conservative, likely due to the fact that the mean of the p-values is sufficiently smaller than $0.5$, and the new estimator the second least conservative. In contrast, for two-sided p-values, the new estimator seems to be the least conservative, and the median based estimator estimator $\hat{\pi}_{0}^{BKY}$ in \cite{Benjamini:2006} the second least conservative but with relatively large variance. Note that $\hat{\pi}_{0}^{BKY}$ can be very accurate when $\pi_0=0.5$ and it is applied to two-sided p-values of FETs, likely due the fact that in this scenario the median of the p-values is close to $0$.
The FDR procedures either have very low power (e.g., when applied to p-values of Fisher's exact tests) or have some power but uncontrolled FDRs (e.g., when applied to p-values of binomial tests), likely due to the bimodality of the empirical CDF's of the p-values mentioned previously. However, the BHH procedure and adaptive BHH procedure are slightly more powerful than the others.

%%%%%%%%%%%%%%%%%%%%%%%%%%%%%%%%%%%%%%%%%%%%%%%%%%%%%%%%%%%%%%%%%%%%%%%%%%%%%
%%%%%%%%%%%%  Application of real data                        %%%%%%%%%%%%
%%%%%%%%%%%%%%%%%%%%%%%%%%%%%%%%%%%%%%%%%%%%%%%%%%%%%%%%%%%%%%%%%%%%%%%%%%%%%
\section{An application to multiple testing with discrete data}\label{sec:app}

We now apply the new estimator and the induced adaptive procedures to multiple testing in a study of HIV vaccine efficacy.
The aim of the study is to identify, among $m=118$ positions, the ``differentially polymorphic'' positions, i.e., the positions where the probability of a non-consensus amino-acid differs between the two amino-acid sequence sets, where the sequence sets were obtained from $n=73$ individuals infected with subtype C HIV (categorized into Group 1) and $n=73$ individuals with subtype B HIV (categorized into Group 2), respectively. Details on how the data were collected and processed can be found in \cite{Gilbert:2005} and references therein.

The multiple testing problem can be stated formally as follows. For each $i = 1,\ldots,118$, let $\theta_{i1}$ and $\theta_{i2}$ respectively be the probabilities of a non-consensus amino-acid at position $i$ for Group 1 and Group 2 sequences. The goal is to test simultaneously the null hypotheses $H_{i0} : \theta_{i1} = \theta_{i2}$ for each $i$, for which the proportion of true null hypotheses is simply $\pi_0 = m^{-1}\vert \left\{i: \theta_{i1} = \theta_{i2}\right\} \vert$. Let $c_{1i}$ and $c_{2i}$ be the number of observed non-consensus amino-acids in the sample from Group 1 and Group 2 respectively. Then $c_{ig}$ for each $i$ and $g \in \left\{1,2\right\}$ can be modelled by a binomial random variable $\mathsf{Bin}\left( \theta_{ig}, n\right)$ with probability of success $\theta_{ig}$ and number of trials $n=73$. Set $c_i = c_{1i} + c_{2i}$ as the total observed count for each position $i = 1,\ldots,118$. For each $i$ conditional on $c_i$ and $n$, Fisher's exact test (FET) can be applied to test $H_{i0}$ and its two-sided p-values $P_i$ can be obtained.

\begin{figure}[t!]
\centering
\includegraphics[width=.7\textwidth]{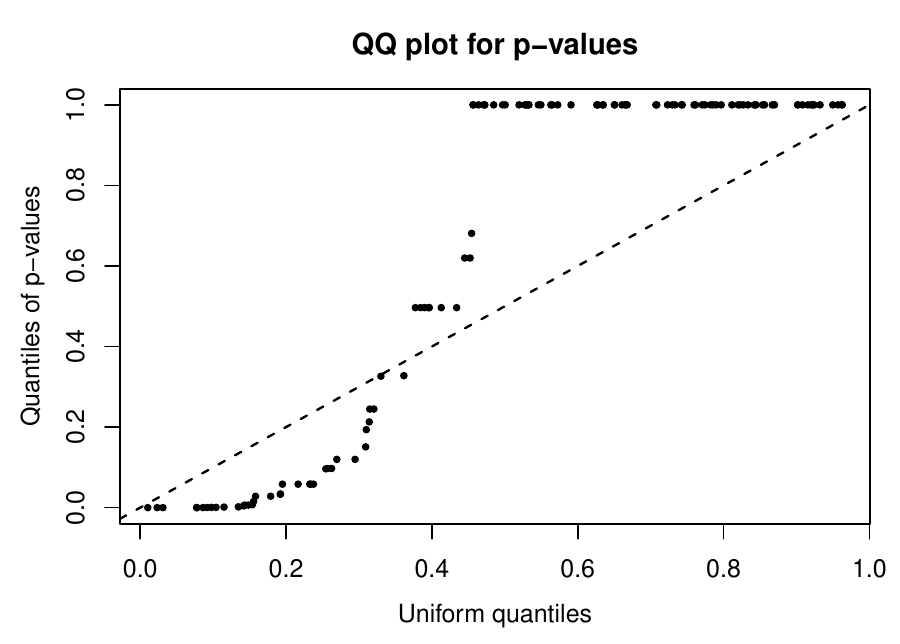}
\caption[QQ plot] {QQ plot of the two-sided p-values of Fisher's exact tests (FETs) in the study of HIV vaccine efficacy. $50$ of these p-values almost surely take value $1$, forming a ``handle'' at height $1$ in the plot.
}
\label{figQQ}
\end{figure}

A summary of the data is provided in Table 1 in \cite{Gilbert:2005}. In particular, there are $50$ positions for which the total observed counts $c_i$ are identically $1$, meaning that the corresponding $50$ two-sided p-values almost surely take value $1$ and their CDF's are Dirac mass at the singleton $\left\{0.5\right\}$. A QQ-plot of the p-values is given by \autoref{figQQ}, where the $50$ p-values that are identically $1$ together form a ``handle'' at height $1$. Based on our findings from the simulations study, these p-values carry too little information about the status of their associated null hypotheses and tend to reduce the power of step-up FDR procedures. Therefore, it may be preferred to exclude these positions from multiple testing. In the following, we will conduct two different analyses of this data set at a nominal FDR level $0.05$.

For the first analysis, all $m=118$ positions are tested simultaneously. \cite{Gilbert:2005} used his modified BH procedure and found $15$ differentially polymorphic positions. The new estimator of $\pi_0$ is $1$, meaning that the induced adaptive procedures reduce to their non-adaptive version. Specifically, the BH procedure found $12$ and the BHH procedure $20$ differentially polymorphic positions.
In the second analysis, we exclude the $50$ positions for which the total observed counts are $1$. The new estimator of $\pi_0$ is $0.7019$. The adaptive BHH procedure found $25$, the BHH procedure $20$, the adaptive BH procedure $16$, and the BH procedure $15$ differentially polymorphic positions, respectively. If the number of observed non-consensus amino-acids are independent, which we tend to believe so, then \autoref{ThmConservABH} on the conservativeness of the adaptive BH procedure suggests that the extra differentially polymorphic positions found by the adaptive BH procedure compared to the BH procedure is worthy of further investigation into its effects on HIV vaccine efficacy.
It is not surprising to observe that, after excluding the $50$ positions whose corresponding p-values almost surely take value $1$, Gilbert's procedure and the BH procedure found the same number, $15$, of differentially polymorphic positions since Gilbert's procedure essentially excluded these same 50 p-values.
However, for this analysis we have not compared the BH and BHH procedure with Gilbert's procedure since the latter is coded in \textsf{Fortran} and \textsf{S}.
In either analysis, the extra differentially polymorphic position found by the (adaptive) BHH procedure are worthy of further investigation in the efficacy study, had we been able to prove the conservativeness of these two procedures.

%%% qq plot of p-values

%%%%%%%%%%%%%%%%%%%%%%%%%%%%%%%%%%%%%%%%%%%%%%%%%%%%%%%%%%%%%
%%%%%%%%%%%%   Discussions                       %%%%%%%%%%%%
%%%%%%%%%%%%%%%%%%%%%%%%%%%%%%%%%%%%%%%%%%%%%%%%%%%%%%%%%%%%%%

\section{Discussion}\label{SecDiscussion}

We have proposed a new estimator of the proportion $\pi_0$ of true null hypotheses for multiple testing in the discrete paradigm, where p-values have discrete and heterogeneous null distributions. It is conservative and less upwardly biased than three popular estimators of the proportion. For multiple testing in the discrete paradigm, the new estimator induces two adaptive FDR procedures, i.e., an adaptive Benjamini-Hochberg procedure that is theoretically proved to be conservative, and an adaptive Benjamini-Hochberg-Heyse (BHH) procedure that is empirically shown to be conservative and more powerful than three other procedures.

The new estimator of $\pi_0$ is designed for discrete p-values whose distributions are heterogeneous. \cite{Liang:2015} developed an estimator $\hat{\pi}_0^L$ of $\pi_0$ for p-values that have discrete but identical distributions, such as those induced by permutation test.
We have compared our estimator with $\hat{\pi}_0^L$ based on p-values of permutation test and found out that our estimator is more conservative than $\hat{\pi}_0^L$. However, the estimator $\hat{\pi}_0^L$ in \cite{Liang:2015} cannot be applied to the simulation settings we have considered where p-values have heterogeneous distributions.
Thus, our estimator and $\hat{\pi}_0^L$ are essentially different and not directly comparable.

The BHH procedure and its adaptive version are empirically shown to be conservative by our simulation study. However, a theoretical justification of the observation is very challenging in the discrete paradigm where null p-values have discrete, heterogeneous distributions. In fact, we do not even have a complete understanding of the threshold sequence implicitly used by the BHH procedure, and the criteria given in \cite{Blanchard:2008} that ensure the conservativeness of an FDR procedure may not be applicable to the BHH procedure. We leave the endeavor along this line to future research.

%%%%%%%%%%%%%%%%%%%%%%%%%%%%%%%%%%%%%%%%%%%%%%%%%%%%%%%%%%
%%%% acknownledgements
%%%%%%%%%%%%%%%%%%%%%%%%%%%%%%%%%%%%%%%%%%%%%%%%%%%%%%%%%%
\section*{Acknowledgements}

We would like to thank Joshua D. Habiger for explaining how to implement his procedure in \cite{Habiger:2015} and Arnold Janssen for providing two references, i.e., \cite{Heesen:2015} and \cite{Heesen:2016} on inequalities for some adaptive FDR procedures.

%\section*{Conflict of interest}
%
%The authors have declared no conflict of interest.

% Bias and std dev: lower tail

% Bias and std dev: two-sided

% FDR and power: lower tail

% TDPs FET: two-sided

%%%%%%%%%%%%%%%%%%%%%%%%%%%%%%%%%%%%%%%%%%%%%%%%%%%%%%%%%%%%
\newpage
\numberwithin{equation}{section}
\numberwithin{figure}{section}

%% begin of appendix
\numberwithin{equation}{section}
\appendix
\section*{Appendices}

We provide in \autoref{appProofs} the proofs of the conservativeness of the new estimator and of the adaptive Benjamini-Hochberg (BH) procedure, in \autoref{secBHHProc} the Benjamini-Hochberg-Heyse (BHH) procedure, in \autoref{secMinsiterp} two misinterpretations of the BHH procedure, and in \autoref{secSimDep} a simulation study on the new estimator and the adaptive BH and adaptive BHH procedures under dependence.

\section{Proofs}\label{appProofs}

\subsection{Proof of \autoref{thm:Conservative}}\label{secConserpi0}

Recall that $I_{0}$ is the set of true null hypotheses and $I_{1}$ that of the false null hypotheses.
Pick any $j$ between $1$ and $n$.
Recall%
\[
\beta\left(\tau_j\right)  =\frac{1}{\left(  1-\tau_{j}\right)  m}%
+\frac{1}{m}\sum\nolimits_{i\in Q_{m}\setminus C}\frac{\mathbf{1}_{\left\{
P_{i}>\lambda_{ij}\right\}  }}{1-\lambda_{ij}}+\frac{1}{m}\left\vert
C\right\vert .
\]
Let $\delta_{j2}=\left(  \left(  1-\tau_{j}\right)  m\right)  ^{-1}$. It is
easy to see%
\begin{align*}
 \mathbb{E}\left( \beta\left(\tau_j\right)  \right)
& =   \delta_{j2}+\frac{1}{m}\sum\nolimits_{i\in I_{0}\setminus C}\frac
{1-F_{i}\left(  \lambda_{ij}\right)  }{1-\lambda_{ij}}+\frac{1}{m}\left\vert
C\right\vert +\frac{1}{m}\sum\nolimits_{i\in I_{1}\setminus C}\frac
{1-G_{i}\left(  \lambda_{ij}\right)  }{1-\lambda_{ij}}\\
& =   \delta_{j2}+\frac{1}{m}\left\vert I_{0}\setminus C\right\vert +\frac
{1}{m}\left\vert C\right\vert +\frac{1}{m}\sum\nolimits_{i\in I_{1}\setminus
C}\frac{1-G_{i}\left(  \lambda_{ij}\right)  }{1-\lambda_{ij}}\\
& =  \delta_{j2}+\frac{1}{m}\left\vert I_{0}\right\vert +\frac{1}{m}%
\sum\nolimits_{i\in I_{1}\setminus C}\frac{1-G_{i}\left(  \lambda_{ij}\right)
}{1-\lambda_{ij}},
\end{align*}
where the second equality follows from $F_{i}\left(  \lambda_{ij}\right)
=\lambda_{ij}$ since $\lambda_{ij}\in S_{i}$. So, the bias%
\[
\delta_{j}=\mathbb{E}\left( \beta\left(\tau_j\right)  -\pi_{0}\right)
=\delta_{j2}+\delta_{j1},
\]
where%
\[
\delta_{j1}=\frac{1}{m}\sum\nolimits_{i\in I_{1}\setminus C}\frac
{1-G_{i}\left(  \lambda_{ij}\right)  }{1-\lambda_{ij}}.
\]
In other words, the bias of $\beta\left(\tau_j\right)$ associated with the null p-values is exactly $0$.
Since $\delta_{j2}>0$ and $\delta_{j1}\geq0$, $\beta\left(\tau_j\right)  $ is conservative.
Since $\hat{\pi}_{0}^{G}=\frac{1}{n}\sum_{j=1}^n \beta\left( \tau _{j}\right)$, the claims hold.

\subsection{Proof of \autoref{ThmConservABH}}\label{secConservABH}

Recall the following: $Q_{s}=\left\{  1,\ldots,s\right\}  $ for each natural number $s$;
$\hat{\pi}_{0}^{G}=n^{-1}\sum_{j=1}^{n}\beta\left(\tau_j\right)  $;
$\mathbf{p}_{0,k}=\left\{  P_{1},\ldots,P_{k-1},0,P_{k+1},\ldots,P_{m}\right\}$
for each $k\in I_{0}$; $\beta_{k}\left(\tau_j\right)  $ is the trial
estimator with guiding value $\tau_{j}$ obtained by applying \autoref{Alg:newest} to $\mathbf{p}_{0,k}$;
$\hat{\pi}_{0,k}^{G}=n^{-1}\sum_{j=1}^{n}\beta_{k}\left(\tau_j\right)$.

Our proof will use the identity provided by the proof of Lemma 1 of \cite{Benjamini:2006} and Theorem 11 of \cite{Blanchard:2009}. In particular, the inequalities \eqref{eqe2A}, i.e., $\mathbb{E}\left( 1/\beta_{k}\left(\tau_j\right) \right) \leq \pi_{0}^{-1}$ and $\mathbb{E}\left(  1/\hat{\pi}_{0,k}^{G} \right)  \leq \pi_{0}^{-1}$, will be proved in the process of proving the conservativeness of the adaptive BH procedure.

Let $\alpha$ be the nominal FDR level. Since the adaptive BH procedure induced
by $\hat{\pi}_{0}^{G}$ is non-increasing and self-consistent with the linear threshold sequence
$\left\{\frac{i\alpha}{m\hat{\pi}_{0}^{G}},1\leq i\leq m \right\}$ (see Definition 3 of \cite{Blanchard:2009} for the non-increasing
and self-consistent property of an FDR procedure) and $1/\hat{\pi}_{0}^{G}$
as an estimator of $\pi_{0}^{-1}$ is non-increasing coordinate-wise in $P_{i}%
$, by Theorem 11 of \cite{Blanchard:2009}, to show that the FDR of the adaptive BH
procedure is bounded $\alpha$, it suffices to show that
\begin{equation}
\mathbb{E}\left(  \frac{1}{\hat{\pi}_{0,k}^{G}}\right)  \leq\frac{1}{\pi
_{0}}\text{ for each }k\in I_{0}.\label{eqe1}%
\end{equation}
By the convexity of the mapping $x\longmapsto\frac{1}{x}$ for $x>0$, we see%
\[
\frac{1}{\hat{\pi}_{0,k}^{G}}=\frac{1}{n^{-1}\sum_{j=1}^{n}\beta
_{k}\left(\tau_j\right)  }\leq\frac{1}{n}\sum\nolimits_{j=1}^{n}\frac
{1}{\beta_{k}\left(\tau_j\right)  }%
\]
and%
\[
\mathbb{E}\left(  \frac{1}{\hat{\pi}_{0,k}^{G}}\right)  \leq\frac{1}{n}%
\sum\nolimits_{j=1}^{n}\mathbb{E}\left(  \frac{1}{\beta_{k}\left(  \tau_j
\right)  }\right)  .
\]
So, it suffices to show%
\begin{equation}
\mathbb{E}\left(  \frac{1}{\beta_{k}\left(\tau_j\right)  }\right)
\leq\frac{1}{\pi_{0}}\text{ for each }j\in Q_{n}.\label{eqe2}%
\end{equation}

Now fix a $j\in Q_{n}=\left\{1,\ldots,n\right\}$. We will split the rest of the proof into two cases: $C=\varnothing$ and $C\neq\varnothing$. We will only provide detailed arguments for the case $C=\varnothing$ since the treatment of the case
$C\neq\varnothing$ is very similar.

\textbf{Case 1}: $C=\varnothing$. In this case,
\[
\beta\left(\tau_j\right)  =\frac{1}{\left(  1-\tau_{j}\right)  m}%
+\frac{1}{m}\sum\nolimits_{i=1}^{m}\frac{\mathbf{1}_{\left\{  P_{i}%
>\lambda_{ij}\right\}  }}{1-\lambda_{ij}}
\]
and
\[
\beta_{k}\left(\tau_j\right)  =\frac{1}{\left(  1-\tau_{j}\right)
m}+\frac{1}{m}\sum\nolimits_{\left\{  i:i\neq k\right\}  }\frac{\mathbf{1}%
_{\left\{  P_{i}>\lambda_{ij}\right\}  }}{1-\lambda_{ij}}.%
\]
Set $y_{ij}=\left(  1-\lambda_{ij}\right)
^{-1}\mathbf{1}_{\left\{  P_{i}>\lambda_{ij}\right\}  }$ for $i\in Q_{m}=\left\{1,\ldots,m\right\}$. Then
$\beta_{k}\left(\tau_j\right)$ can be rewritten as%
\[
\beta_{k}\left(\tau_j\right)  =\frac{1}{\left(  1-\tau_{j}\right)
m}+\frac{1}{m}\sum\nolimits_{\left\{  i:i\neq k\right\}  }y_{ij}.
\]
So,
\begin{equation}
m\beta_{k}\left(\tau_j\right)  \geq\left(  1-\tau_{j}\right)  ^{-1}%
+\sum\nolimits_{i\in I_{0}\backslash\left\{  k\right\}  }y_{ij}.\label{eqe14}%
\end{equation}
and
\begin{equation}
\mathbb{E}\left(  \frac{1}{\beta_{k}\left(  \tau_j\right)  }\right)  \leq
m\mathbb{E}\left(  \frac{1}{\left(  1-\tau_{j}\right)  ^{-1}+\sum
\nolimits_{i\in I_{0}\backslash\left\{  k\right\}  }y_{ij}}\right)
.\label{eqe16}%
\end{equation}

Recall $y_{ij}=\left(  1-\lambda_{ij}\right)  ^{-1}\mathbf{1}_{\left\{
P_{i}>\lambda_{ij}\right\}  }$. Since $\lambda_{ij}\in S_{i}$ and $S_{i}$ is
the support of p-value $P_{i}$, we have%
\begin{equation}
\left\{
\begin{array}
[c]{c}%
F_{i}\left(  \lambda_{ij}\right)  =\lambda_{ij},i\in I_{0}\\
\mathbb{P}\left(  y_{ij}=0\right)  =\lambda_{ij},i\in I_{0}\\
\mathbb{P}\left(  y_{ij}=\frac{1}{1-\lambda_{ij}}\right)  =1-\lambda_{ij},i\in
I_{0}%
\end{array}
\right.  \label{eqe18}%
\end{equation}
Since $\max_{i\in Q_{m}}\lambda_{ij}\leq\tau_{j}$, the identity
(\ref{eqe18}) implies $y_{ij}\geq\frac{1}{1-\tau_{j}}$ when $y_{ij}\neq0$. Set
$w_{ij}=\mathbf{1}_{\left\{  P_{i}>\lambda_{ij}\right\}  }$. Then%
\[
\left\{
\begin{array}
[c]{c}%
y_{ij}=0\text{ if and only if }w_{ij}=0\\
y_{ij}\geq\frac{1}{1-\tau_{j}}>w_{ij}=1\text{ \ whenever }y_{ij}\neq0
\end{array}
\right.
\]
and%
\begin{equation}
\sum\nolimits_{i\in I_{0}\backslash\left\{  k\right\}  }y_{ij}\geq\frac
{1}{1-\tau_{j}}\sum\nolimits_{i\in I_{0}\backslash\left\{  k\right\}  }%
w_{ij}.\label{eqe19}%
\end{equation}
So, setting $W_{j,k}=\sum\nolimits_{i\in I_{0}\backslash\{k\}}w_{ij}$ gives
\begin{equation}
\mathbb{E}\left(  \frac{1}{\left(  1-\tau_{j}\right)  ^{-1}+\sum
\nolimits_{i\in I_{0}\backslash\left\{  k\right\}  }y_{ij}}\right)
\leq\left(  1-\tau_{j}\right)  \mathbb{E}\left(  \frac{1}{1+W_{j,k}}\right)
.\label{eqe20}%
\end{equation}

On the other hand, setting $\tilde{w}_{ij}=\mathbf{1}_{\left\{  P_{i}%
>\tau_{j}\right\}  }$ gives $w_{ij}\ge\tilde{w}_{ij}$ again due to $\max_{i\in Q_{m}%
}\lambda_{ij}\leq\tau_{j}$. Let $\tilde{W}_{j,k}=\sum\nolimits_{i\in
I_{0}\backslash\{k\}}\tilde{w}_{ij}$. Then $\tilde{W}_{j,k}$ is a Binomial
random variable with probability of success $1-\tau_{j}$ and total number of
trials $m_{0}-1$. Further,
\begin{equation}
\mathbb{E}\left(  \frac{1}{1+W_{j,k}}\right)  \leq\mathbb{E}\left(  \frac
{1}{1+\tilde{W}_{j,k}}\right)  =\frac{1-\tau_{j}^{m_{0}}}{m_{0}\left(
1-\tau_{j}\right)  },\label{eqe21}%
\end{equation}
where the equality has been derived from the identity provided in the proof of
Lemma 1 of \cite{Benjamini:2006}. Combining (\ref{eqe16}), (\ref{eqe20}) and
(\ref{eqe21}), we obtain%
\begin{align}
\mathbb{E}\left(  \frac{1}{\beta_{k}\left(  \tau_j\right)  }\right)   &
\leq\left(  1-\tau_{j}\right)  \frac{m}{m_{0}}\frac{1-\tau_{j}^{m_{0}}}%
{1-\tau_{j}}=\frac{m}{m_{0}}\left(  1-\tau_{j}^{m_{0}}\right)  \label{eqe25}\\
&  < \frac{m}{m_{0}}=\frac{1}{\pi_{0}}.\nonumber
\end{align}
Namely, (\ref{eqe2}) holds, and so does (\ref{eqe1}), i.e., \eqref{eqe2A} holds. Therefore, the adaptive BH
procedure is conservative.

\textbf{Case 2}:\ $C\neq\varnothing$. In this case,%
\[
\beta\left(\tau_j\right)  =\frac{1}{\left(  1-\tau_{j}\right)  m}%
+\frac{1}{m}\sum\nolimits_{i\in Q_{m}\setminus C}\frac{\mathbf{1}_{\left\{
P_{i}>\lambda_{ij}\right\}  }}{1-\lambda_{ij}}+\frac{1}{m}\left\vert
C\right\vert .
\]
Since $\left\vert C\right\vert \geq1$ and $I_{0}\subseteq Q_{m}\setminus C$,
we see
\begin{equation}
\beta\left(\tau_j\right)  \geq\frac{1}{\left(  1-\tau_{j}\right)
m}+\frac{1}{m}\sum\nolimits_{i\in I_{0}}\frac{\mathbf{1}_{\left\{
P_{i}>\lambda_{ij}\right\}  }}{1-\lambda_{ij}}+\frac{1}{m}\label{eqe26}%
\end{equation}
and
\begin{equation}
\beta_{k}\left(\tau_j\right)  \geq\frac{1}{\left(  1-\tau_{j}\right)
m}+\frac{1}{m}\sum\nolimits_{i\in I_{0} \setminus {\left\{k\right\}}}\frac{\mathbf{1}_{\left\{
P_{i}>\lambda_{ij}\right\}  }}{1-\lambda_{ij}}+\frac{1}{m}.\label{eqe26a}%
\end{equation}
Applying the arguments for the case $C=\varnothing$ directly leads to (\ref{eqe14}),
(\ref{eqe16}), (\ref{eqe20}) and (\ref{eqe21}) and (\ref{eqe25}). So,
(\ref{eqe2}) and (\ref{eqe1}) hold, and the adaptive BH procedure is
conservative.

\section{The Benjamini-Hochberg-Heyse Procedure}\label{secBHHProc}

Let $\left\{  P_{i}\right\}  _{i=1}^{m}$ be p-values such that under the true
null hypothesis $\mathbb{P}\left(  P_{i}\leq t\right)  \leq t$ for $t\in\left[
0,1\right]  $. For each $1\leq i\leq m$, let $p_{i}$ be the observed value of
$P_{i}$, $H_{i}$ be the null hypothesis associated with $p_{i}$, $\left\{
p_{\left(  i\right)  }\right\}  _{i=1}^{m}$ the order statistics of $\left\{
p_{i}\right\}  _{i=1}^{m}$ such that $p_{\left(  1\right)  }\leq p_{\left(
2\right)  }\leq\cdots\leq p_{\left(  m\right)  }$, and $H_{\left(  i\right)
}$ the null hypothesis associated with $p_{\left(  i\right)  }$.

The Benjamin-Hochberg (BH) procedure of \cite{Benjamini:1995} sets%
\begin{equation}
\theta=\max\left\{  i:p_{\left(  i\right)  }\leq\frac{i}{m}\alpha\right\}
\label{eqThreshBH}%
\end{equation}
and rejects $H_{\left(  j\right)  }$ for $1\leq j\leq\theta$ if $\theta$
exits. In \cite{Heyse:2011} the BH procedure is equivalently rephrased as follows: let $p_{\left[
m\right]  }=p_{\left(  m\right)  }$,%
\begin{equation}
p_{\left[  i\right]  }=\min\left\{  p_{\left[  i+1\right]  },\frac{mp_{\left(
i\right)  }}{i}\right\}  \text{ for }1\leq i\leq m-1 \label{eqEqiSeq}%
\end{equation}
and%
\begin{equation}
\varepsilon=\max\left\{  i:p_{\left[  j\right]  }\leq\alpha\right\}  ;
\label{eqEqiSeq2}%
\end{equation}
then reject all $H_{\left(  j\right)  }$ for which $j\leq\varepsilon$ if
$\varepsilon$ exits.

In order to account for the discreteness of p-value distributions, \cite{Heyse:2011}
proposed the ``Benjamini-Hochberg-Heyse (BHH)'' procedure, a
modification and extension of the BH procedure, that is empirically shown to be conservative
and more powerful than the BH procedure for multiple testing based on discrete p-values.
For each $1\leq j\leq
m$ and $p\in\left[  0,1\right]  $, let $g_{j}\left(  p\right)  $ be the
largest value achievable by $P_{j}$ that is less than or equal to $p$, for
which $g_{j}\left(  p\right)  =0$ is set if the smallest value achievable by
$P_{j}$ is larger than $p$. Define
\begin{equation}
Q\left(  p_{\left(i\right)}\right)  =\sum_{j=1}^{m}g_{j}\left(  p_{\left(  i\right)
}\right)  \text{ \ for \ }i=1,\ldots,m. \label{eqHseq3}%
\end{equation}
The BHH procedure is defined as follows. Let $p_{\left\langle m\right\rangle
}=p_{\left(  m\right)  }$,%
\begin{equation}
p_{\left\langle i\right\rangle }=\min\left\{  p_{\left[  i+1\right]  }%
,i^{-1}Q\left(  p_{\left(i\right)}\right) \right\}  \text{ for }1\leq i\leq m-1
\label{eqHseq}%
\end{equation}
and%
\begin{equation}
\eta=\max\left\{  i:p_{\left\langle j\right\rangle }\leq\alpha\right\}  ;
\label{eqHeq2}%
\end{equation}
then reject all $H_{\left(  j\right)  }$ for which $j\leq\eta$ if $\eta$
exits.
It is important to note that the BHH procedure accounts for the step-up sequence induced by the BH procedure;
see \eqref{eqHseq1} of \autoref{lmRephrase} for the expression for $p_{\left\langle i\right\rangle }$.

We have the following result:
\begin{lemma}
\label{lmRephrase}The following hold:

\begin{enumerate}
\item $p_{\left[  m\right]  }=p_{\left(  m\right)  }=p_{\left\langle
m\right\rangle }$. For $1\leq i\leq m-1$, $p_{\left[  i\right]  }\leq
p_{\left[  i+1\right]  }$, $p_{\left\langle i\right\rangle }\leq
p_{\left\langle i+1\right\rangle }$, $p_{\left[  i\right]  }\geq
p_{\left\langle i\right\rangle }$ and $Q\left(  p_{\left(  i\right)  }\right)
\leq Q\left(  p_{\left(  i+1\right)  }\right)  $. If all p-values have
continuous distributions, then $p_{\left\langle i\right\rangle }=p_{\left[
i\right]  }$ for all $1\leq i\leq m$.

\item For any $1\leq s\leq m-1$,%
\begin{equation}
p_{\left[  m-s\right]  }=\min\left\{  p_{\left(  m\right)  },\frac{mp_{\left(
m-1\right)  }}{m-1},\ldots,\frac{mp_{\left(  m-s+1\right)  }}{m-s+1}%
,\frac{mp_{\left(  m-s\right)  }}{m-s}\right\}  .\label{eqEqiSeq1}%
\end{equation}

\item For any $1\leq s\leq m-1$,%
\begin{equation}
p_{\left\langle m-s\right\rangle }=\min\left\{  p_{\left(  m\right)  }%
,\frac{mp_{\left(  m-1\right)  }}{m-1},\ldots,\frac{mp_{\left(  m-s+1\right)
}}{m-s+1},\frac{Q\left(p_{\left(  m-s\right)  }\right)}{m-s}\right\}  .\label{eqHseq1}%
\end{equation}

\item The BH procedure and its rephrased version are equivalent, i.e., they
always reject the same set of null hypotheses.
\end{enumerate}
\end{lemma}

\begin{proof}
The first claim is obvious. By the definition in (\ref{eqEqiSeq}), we see%
\[
p_{\left[  m-1\right]  }=\min\left\{  p_{\left(  m\right)  },\frac{mp_{\left(
m-1\right)  }}{m-1}\right\}  .
\]
By mathematical induction, we obtain (\ref{eqEqiSeq1}) for any $1\leq s\leq
m-1$. By the definition in (\ref{eqHseq}), we see%
\[
p_{\left\langle m\right\rangle }=\min\left\{  p_{\left(  m\right)  },\left(
m-1\right)  ^{-1}Q\left(p_{\left(  m-1\right)  }\right)\right\}  .
\]
Using (\ref{eqHseq}) and (\ref{eqEqiSeq1}), we obtain (\ref{eqHseq1}) for any
$1\leq s\leq m-1$. The two quantities $p_{\left[  i\right]  }$ and
$p_{\left\langle i\right\rangle }$ differ by the last element from which the
minima are taken.

Now we show the equivalence between the BH procedure and its rephrased
version. Recall the indices defined in (\ref{eqThreshBH}) and (\ref{eqEqiSeq2}%
). $\theta$ does not exist if and only if $p_{\left(  i\right)  }>\frac{i}%
{m}\alpha$ \ for all $1\leq i\leq m$ if and only if $p_{\left[  i\right]
}>\alpha$ for\ all $1\leq i\leq m$ if and only if $\varepsilon$ does not exit.
In other words, neither procedures make any rejections or both make some
rejections. Therefore, it is left to show $\theta=\varepsilon$ when either
$\theta$ or $\varepsilon$ exists.

Fix some index $l$ between $1$ and $m$. Then, $p_{\left(  l\right)  }\leq
\frac{l}{m}\alpha$ and $\frac{mp_{\left(  j\right)  }}{j}>\alpha$ for all
$j>l$ if and only if $p_{\left[  l\right]  }=\frac{mp_{\left(  l\right)  }}%
{l}$ by (\ref{eqHseq1}), $p_{\left[  l\right]  }\leq\alpha$ and $p_{\left[
j\right]  }>\alpha$ for all $j>l$. However, $p_{\left[  i\right]  }$ is
nondecreasing in $i$ for $1\leq i\leq m$. Therefore, $\theta=\varepsilon$.
This completes the proof.
\end{proof}

\section{Two misinterpretations of the BHH procedure}\label{secMinsiterp}

In this section, we point out two misinterpretations of the the Benjamini-Hochberg-Heyse (BHH) procedure, one from
\cite{Heller:2012} and the other from \cite{DOHLER2016}.

Section 2.2 of \cite{Heller:2012} mistakenly rephrased the BHH procedure as follows: let
\begin{equation}
p_{\left\{  i\right\}  }=\min_{j\geq i}\frac{\sum_{j=1}^{m}\Pr\left(
P_{j}\leq p_{\left(  i\right)  }\right)  }{j} \label{eqHeller1}%
\end{equation}
and reject $H_{\left(  i\right)  }$ if $i$ is such that $p_{\left\{
i\right\}  }\leq\alpha$.
Clearly, $p_{\left\{
m\right\}  }=\frac{Q\left(  p_{\left(  m\right)  }\right)  }{m}$ and
$p_{\left\{  m\right\}  }\leq p_{\left(  m\right)  }$. Further,%
\begin{equation}
p_{\left\{  i\right\}  }=\min_{j\geq i}\frac{Q\left(  p_{\left(  j\right)
}\right)  }{j}=\min\left\{  \frac{Q\left(  p_{\left(  m\right)  }\right)  }%
{m},\ldots,\frac{Q\left(  p_{\left(  i\right)  }\right)  }{i}\right\}
,\label{eqH2}%
\end{equation}
and $p_{\left\{  i\right\}  }\leq p_{\left\{  i+1\right\}  }$ for $1\leq i\leq
m-1$. So, $p_{\left\{  i\right\}  }$ is not almost surely equal to%
\[
p_{\left\langle i\right\rangle }=\min\left\{  p_{\left(  m\right)  }%
,\frac{mp_{\left(  m-1\right)  }}{m-1},\ldots,\frac{mp_{\left(  i+1\right)  }%
}{i+1},\frac{Q\left(  p_{\left(  i\right)  }\right)  }{i}\right\}
\]
for all $1\leq i\leq m-1$; see \eqref{eqHseq1} in \autoref{lmRephrase} for the expression for $p_{\left\langle i \right\rangle }$.
Namely, the rephrased version does not account for the step-up sequence induced by the BH procedure
and is not equivalent to the BHH procedure.
In particular, we have the following.
Let $\xi=\max\left\{i:p_{\left\{  i\right\}  }\leq\alpha\right\}  $. Then the rephrased procedure
is equivalent to rejecting $H_{\left(  i\right)  }$ if $i\leq\xi$ when $\xi$
exists. If $\frac{Q\left(  p_{\left(  j\right)  }\right)  }{j}>\alpha$ for all
$1\leq j\leq m$, then $p_{\left\langle j\right\rangle }>\alpha$ for all $1\leq
j\leq m$. However, even though $p_{\left\langle j\right\rangle }>\alpha$ for
all $1\leq j\leq m$, $p_{\left\{  m\right\}  }<\alpha$ can happen when%
\begin{equation}
\frac{Q\left(  p_{\left(  m\right)  }\right)  }{m}<\alpha<p_{\left(  m\right)
}. \label{eqNoneq}%
\end{equation}
Namely, the rephrased version is not equivalent to the BHH procedure when the latter
rejects all null hypotheses.

Appendix 1 of \cite{DOHLER2016} mistakenly rephrased the BHH procedure as follows:
let $\tilde{p}_{\left(  m\right)  }=p_{\left(  m\right)  }$ and%
\begin{equation}\label{eqc1}
\tilde{p}_{\left(  i\right)  }=\min\left\{  \tilde{p}_{\left(  i+1\right)
}, i^{-1} Q\left(  p_{\left(  i\right)  }\right) \right\}
\end{equation}
for $i=1,\ldots,m-1$; reject reject $H_{\left(  i\right)  }$ if $i$ is such
that $\tilde{p}_{\left(  i\right)  }\leq\alpha$. Obviously, this rephrased
procedure is not equivalent to the BHH procedure, since the BHH procedure
contains a modified step-up sequence induced by the BH procedure (see \eqref{eqEqiSeq}, \eqref{eqHseq} and \autoref{lmRephrase})
which is missing from \eqref{eqc1}.
Specifically,%
\begin{equation}
\tilde{p}_{\left(  m-s\right)  }=\min\left\{  p_{\left(  m\right)  }%
,\frac{Q\left(  p_{\left(  m-1\right)  }\right)  }{m-1},\ldots,\frac{Q\left(
p_{\left(  m-s+1\right)  }\right)  }{m-s+1},\frac{Q\left(  p_{\left(
m-s\right)  }\right)  }{m-s}\right\}  \label{eqd1}%
\end{equation}
for any $1\leq s\leq m-1$. Clearly, $\tilde{p}_{\left(  m\right)
}=p_{\left\langle m\right\rangle }$ and $\tilde{p}_{\left(  m-1\right)
}=p_{\left\langle m-1\right\rangle }$. However, $\tilde
{p}_{\left(  m-s\right)  }$ is not almost surely equal to%
\[
p_{\left\langle m-s\right\rangle }=\min\left\{  p_{\left(  m\right)  }%
,\frac{mp_{\left(  m-1\right)  }}{m-1},\ldots,\frac{mp_{\left(  m-s+1\right)
}}{m-s+1},\frac{Q\left(  p_{\left(  m-s\right)  }\right)  }{m-s}\right\}
\]
for all $s=2,\ldots,m-1$; see \eqref{eqHseq1} of \autoref{lmRephrase} for the above expression for $p_{\left\langle m-s\right\rangle }$.

Since the rephrased versions of the BHH procedure in \cite{Heller:2012} and \cite{DOHLER2016} are not equivalent to the BHH procedure, the counterexamples to the rephrased versions given by these articles, where all null hypotheses are true and the FDR is equal to the familywise error rate (FWER), cannot be regarded as counterexamples to the conservativeness of the BHH procedure.

Finally, \cite{Dohler:2017} rephrases the BHH procedure (see equation (8) there) as follows.
Let $\bar{F}\left(  t\right)  =m^{-1}\sum_{i=1}^{m}F_{i}\left(
t\right)  $ for $t\in\left[  0,1\right]  $, where $F_{i}$ is the CDF of
$p_{i}$ obtained by assuming $H_i$ is true, and let $\mathcal{A}$ be the union of all supports of $F_{i}$,
$i=1,\ldots m.$ Reject $H_{i}$ if $p_{i}\leq\tau_{\hat{k}}$, where%
\[
\hat{k}=\max\left\{  k\in\left\{  1,\ldots,m\right\}  :p_{\left(  k\right)
}\leq\tau_{k}\right\}
\]
and%
\[
\tau_{k}=\max\left\{  \tau\in\mathcal{A}:\bar{F}\left(  t\right)  \leq
\frac{\alpha k}{m}\right\}  \text{ \ for \ }1\leq k\leq m.
\]
However, this rephrased version does not account for the step-up sequence induced by the BH procedure and is equivalent to the rephrased version in \cite{Heller:2012} (see \eqref{eqHeller1}), meaning that it is not equivalent to the BHH procedure.

\section{Simulation study under dependence} \label{secSimDep}

We will generate vectors of binomial or Poisson random variables whose correlation matrices approximately are block diagonal and have nonnegative entries, since block diagonal correlation matrix is plausible when functional groups exist in data from various sources.
In other words, the generated binomial or Poisson random variables approximately have positive, blockwise correlation.
The estimators and FDR procedures we compare are the same as those given in \autoref{SecSimuStudy}. However, to ensure that the FDR procedures have moderate power under such dependency, we generate data as follows:

\begin{enumerate}
  \item Set $m=1000$. Let $\mathbf{D} = \mathsf{diag}\left\{\mathbf{D}_1,\mathbf{D}_2,\mathbf{D}_3,\mathbf{D}_4,\mathbf{D}_5\right\}$ be a block diagonal matrix
   with $5$ blocks each of size $200 \times 200$, such that the main diagonal entries of $\mathbf{D}$ are identically $1$ and the off-diagonal entries of
   $\mathbf{D}_i$ are identically $0.1\times i$ for $i=1,\ldots,5$. In other words, $\mathbf{D}$ is a block diagonal correlation matrix such that
   each of its blocks represents a random vector whose entries are equally correlated.
    There is no specific reason for choosing $5$ blocks.
  \item Generate a realization $\mathbf{z}=(z_1,\ldots,z_m)$ from the $m$-dimensional Normal random vector with zero mean and correlation matrix $\mathbf{D}$, and obtain
  the vector $\mathbf{u}=(u_1,\ldots,,u_m)$ of lower-tail probabilities such that $u_i = \Phi(z_i)$, where $\Phi$ is
  the CDF of the standard Normal random variable.
  \item Generate $m$-dimensional vectors of binomial or Poisson
  random variables using the vector $\mathbf{u}$ as the quantiles of the corresponding marginal binomial or Poisson distribution as follows:

  \begin{enumerate}
\item Poisson data: Generate $m$ $\theta _{i1}$'s independently from
the Pareto distribution $\mathsf{Pareto}\left( 5,8\right) $ with location parameter $5$ and shape parameter $5$. Generate $m_{1}$ $\rho _{i}$'s independently from $\mathsf{Unif}\left(
2.5,4.5\right) $. Set $\theta _{i2}=\theta _{i1}$ for $1\leq i\leq m_{0}$ but $%
\theta _{i2}=\rho _{i}\theta _{i1}$ for $m_{0}+1\leq i\leq m$. For each $1 \leq i \leq m$ and $g \in \cbk{1,2}$,
generate a count $\xi _{ig}$ from the Poisson distribution $\mathsf{Poisson}\left( \theta _{ig}\right) $
whose quantile is $u_i$.

\item Binomial data: generate $\theta_{i1}$ from $\mathsf{Unif}\left(
0.15,0.2\right)  $ for $i =1,\ldots,m_{0}$ and set $\theta_{i2}=\theta_{i1}$ for $i =1,\ldots,m_{0}$. Set $\theta_{i1}=0.2$ and
$\theta_{i2}=0.5$ for $i =m_{0}+1,\ldots,m$. Set $n=30$, and for each $g \in \{1,2\}$ and $i$, generate a count $\xi _{ig}$ from the binomial distribution $\mathsf{Bin}\left( \theta_{ig}, n\right)$ whose quantile is $u_i$.

\end{enumerate}

\item Maintain other settings in \autoref{SecSimSteup} in the main paper;
\end{enumerate}

\begin{figure}[htp]
\centering
\includegraphics[height=0.73\textheight,width=1\textwidth]{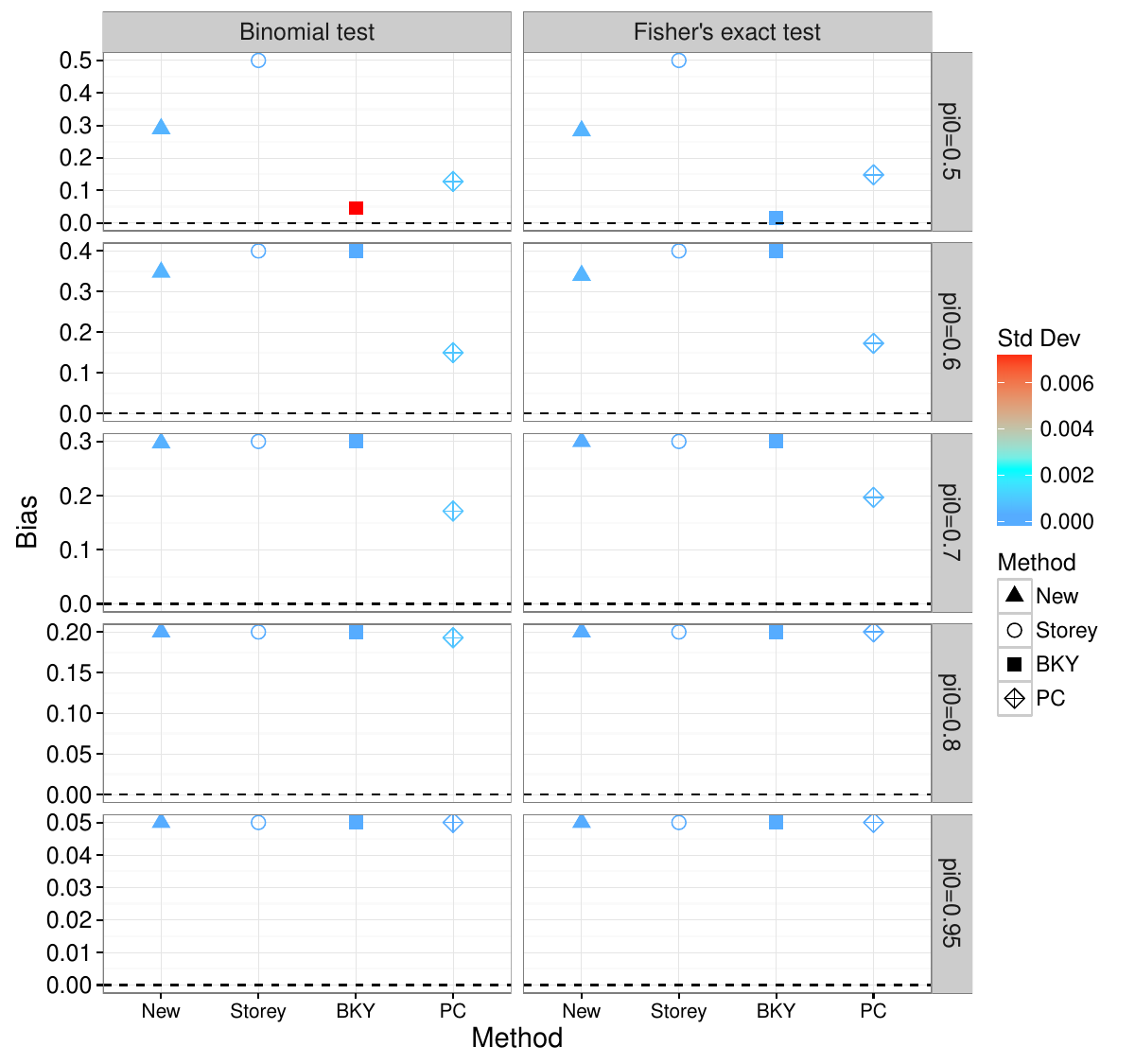}
\caption[] {
Bias and standard deviation (indicated by the color legend ``Std Dev") of each estimator of the proportion $\pi_0$ of true null hypotheses. All estimators have been applied to one-sided p-values of a type of test indicated by the horizontal strip name. The dashed line marks zero bias; ``pi0'' the vertical strip names refers to $\pi _{0}$.
An estimator of $\pi_0$ is said to be better if it has smaller non-negative bias and small standard deviation.
All estimators are very conservative.
The estimator in \cite{Pounds:2006} (indicated by ``PC'' and the diamond) is overall the best and the new estimator (indicated by ``New'' and the triangle) the second best.
}
\label{figbiasstdLTDep}
\end{figure}

\begin{figure}[htp]
\centering
\includegraphics[height=0.73\textheight,width=1\textwidth]{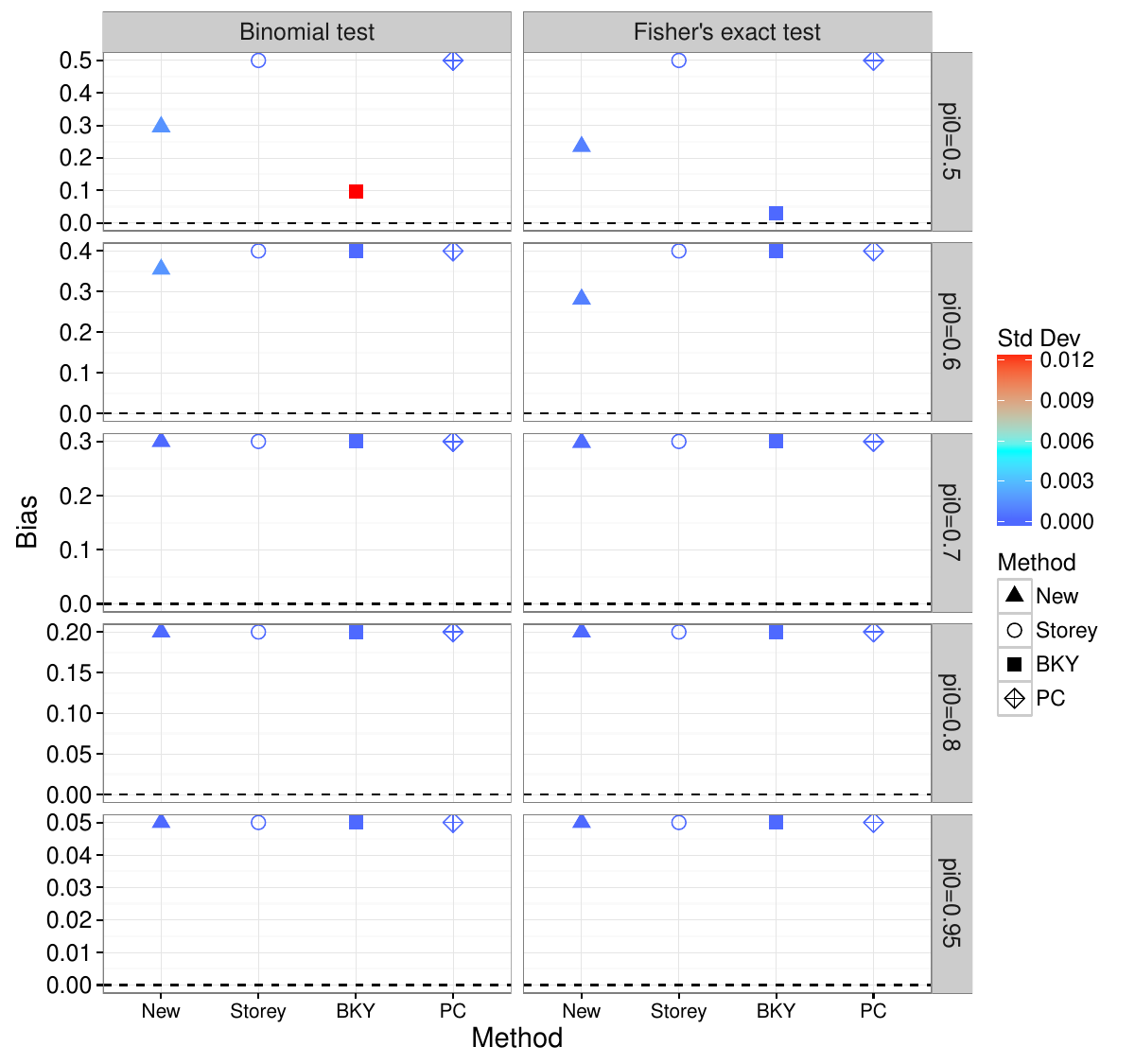}
\caption[] {Bias and standard deviation (indicated by the color legend ``Std Dev") of each estimator of the proportion $\pi_0$ of true null hypotheses. All estimators have been applied to two-sided p-values of a type of test indicated by the horizontal strip name. The dashed line marks zero bias; ``pi0'' the vertical strip names refers to $\pi _{0}$.
An estimator of $\pi_0$ is said to be better if it has smaller non-negative bias and small standard deviation.
All estimators are very conservative. The new estimator (indicated by ``New'' and the triangle) is the best overall and the estimator in \cite{Benjamini:2006} (indicated by ``BKY'' and the square) the second best.
}
\label{figbiasstdDep}
\end{figure}

\begin{figure}[htp]
\centering
\includegraphics[height=0.73\textheight,width=1\textwidth]{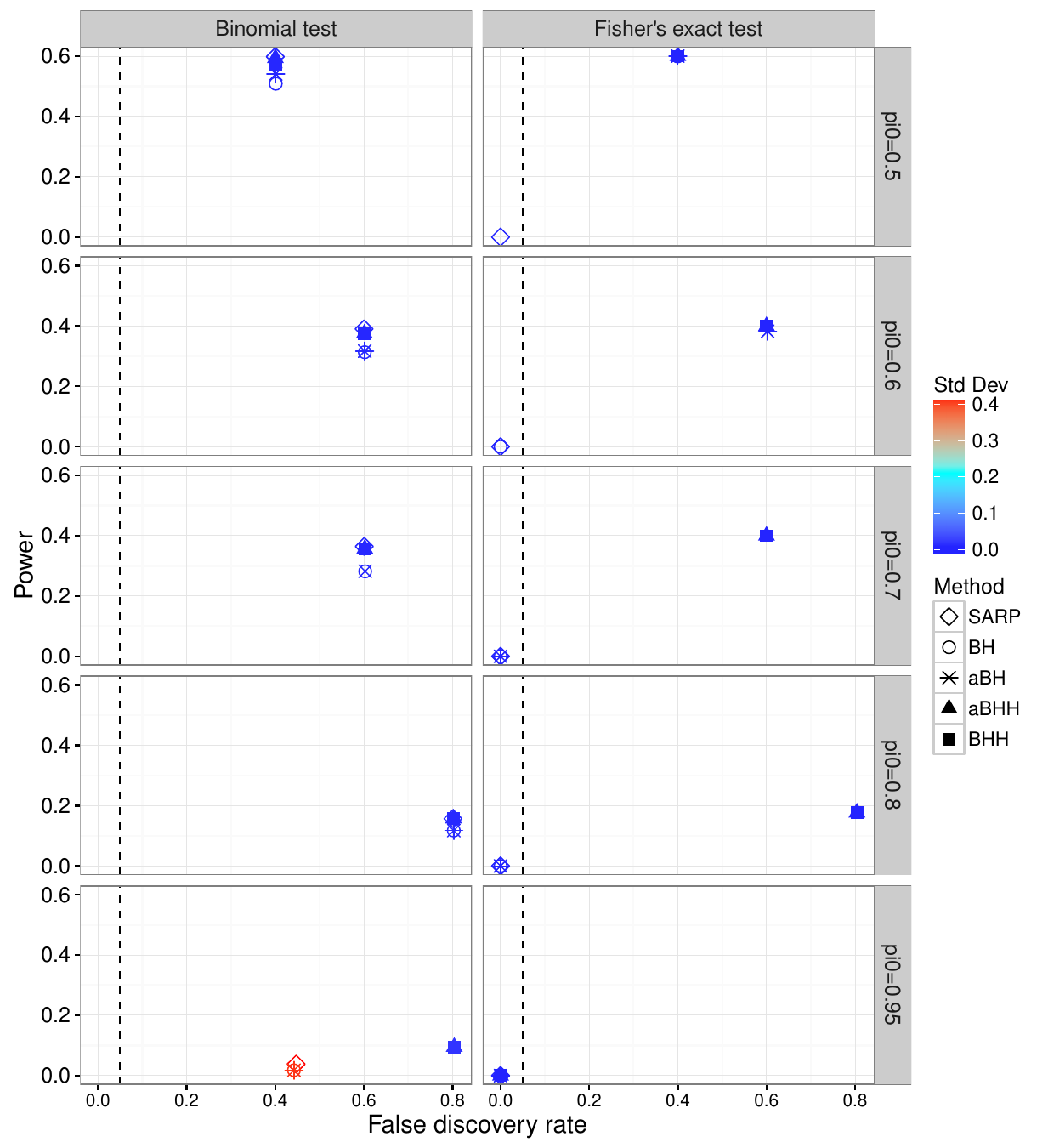}
\caption[] {
False discovery rate (FDR) and power of the competing FDR procedures when they are applied to one-sided p-values of a type of test indicated by the horizontal strip name. In the vertical strip names, ``pi0'' refers to $\pi _{0}$; the color gradient is the standard deviation (Std Dev) of the false discovery proportion whose expectation is the FDR.
All FDR procedures have very low power (e.g., when applied to p-values of Fisher's exact tests) or have some power but uncontrolled FDRs (e.g., when applied to p-values of binomial tests). However, the adaptive BHH procedure ``aBHH", indicated by solid triangle, is overall slightly more powerful than other procedures.}
\label{figPwrLowerDep}
\end{figure}

\begin{figure}[htp]
\centering
\includegraphics[height=0.73\textheight,width=1\textwidth]{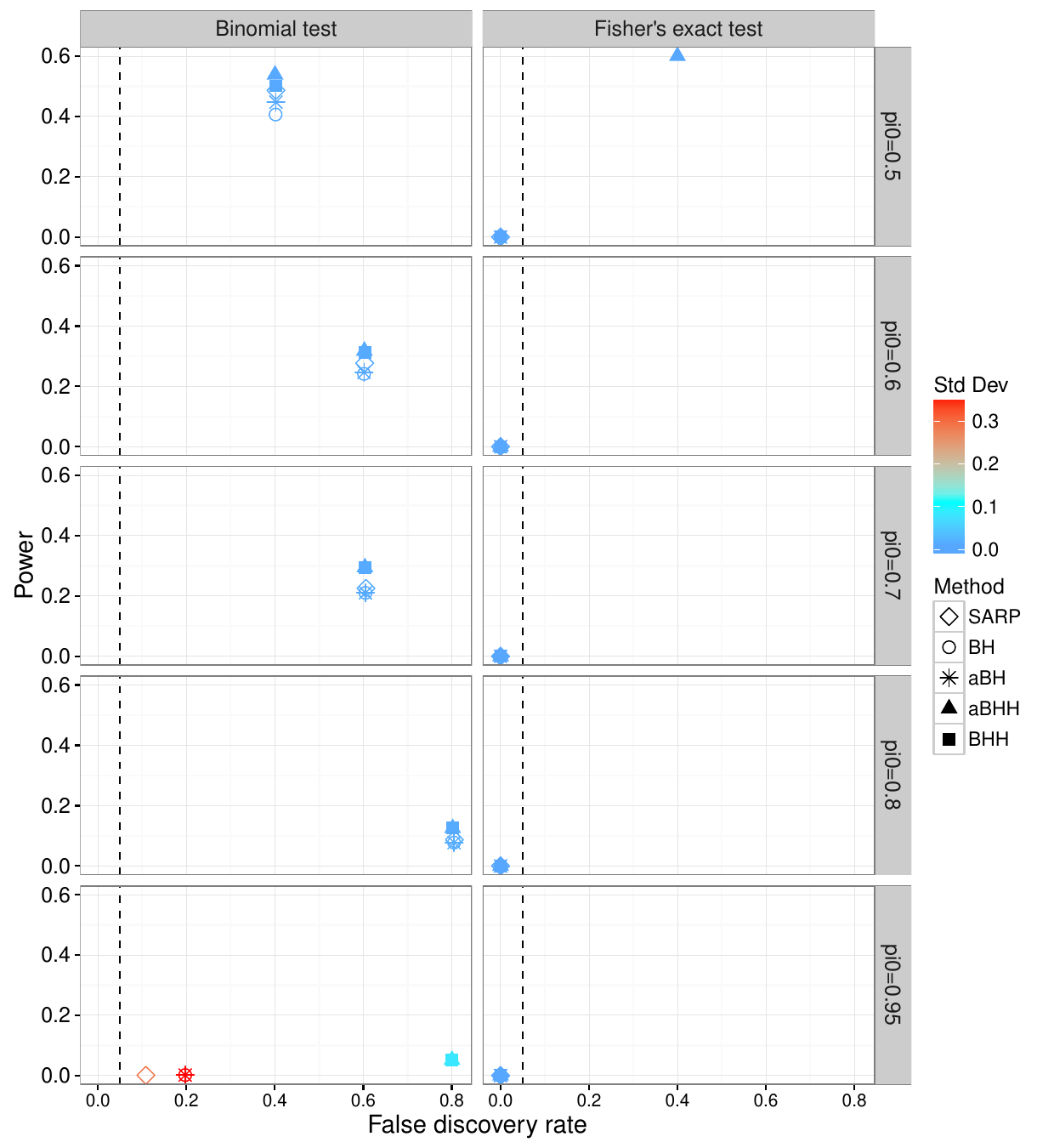}
\caption[FDR and power, two-sided] {False discovery rate (FDR) and power of the competing FDR procedures when they are applied to two-sided p-values of a type of test indicated by the horizontal strip name. In the vertical strip names ``pi0'' refers to $\pi _{0}$; the color gradient is the standard deviation (Std Dev) of the false discovery proportion whose expectation is the FDR.
All FDR procedures have very low power (e.g., when applied to p-values of Fisher's exact tests) or have some power but uncontrolled FDRs (e.g., when applied to p-values of binomial tests). However, the adaptive BHH procedure ``aBHH", indicated by solid triangle, is overall slightly more powerful than other procedures.}
\label{figPwrTwosideDep}
\end{figure}

The simulation results are presented in \autoref{figbiasstdLTDep}, \autoref{figbiasstdDep}, \autoref{figPwrLowerDep} and \autoref{figPwrTwosideDep}. A discussion on the simulation results for this setting is provided at the end of \autoref{secSimRes}.

% Bias and std dev: dep lower tail

% Bias and std dev: two-sided dep

% FDR and power: lower tail dep

% TDPs FET: two-sided dep

%%%%%%%%%%%%%%%%%%%%%%%%%%%%%%%%%%%%%%%%%

\end{document}